\documentclass[11pt,aps,prc,showpacs,
a4paper,preprintnumbers,
superscriptaddress,
nofootinbib]{revtex4-1}

\usepackage[T1]{fontenc}
\usepackage{charter}

\usepackage{graphicx}
\usepackage{epsfig}
\usepackage{amsmath,amsfonts}
\usepackage{mathrsfs}
\usepackage{xcolor}
\usepackage{slashed}
\usepackage[colorlinks,citecolor=blue,linktoc=all,linkcolor=cyan,urlcolor=blue]{hyperref}
\def\beq{\begin{equation}}
\def\eeq{\end{equation}}
\def\bea{\begin{eqnarray}}
\def\eea{\end{eqnarray}}
\def\beqa{\begin{equation}\begin{array}{l}}
\def\eeqa{\end{array}\end{equation}}
\def\barr{\left(\begin{array}{c}}
\def\earr{\end{array}\right)}
\def\bmat{\left(\begin{array}{cc}}
\def\emat{\end{array}\right)}
\def\3d{3-D}

\newcommand{\mat}[1]{\ensuremath \mathsf{#1}}
\newcommand{\svec}[1]{\ensuremath \boldsymbol{#1}}
\newcommand{\svect}[1]{\ensuremath \boldsymbol{#1}^\mathsf{T}}

\newcommand{\bra}[1]{\left\langle #1\right|}
\newcommand{\ket}[1]{\left| #1\right\rangle}
\newcommand{\brkt}[2]{\left\langle #1\vphantom{#2}\right|\left.\vphantom{#1}#2 \right\rangle}
\newcommand{\Y}{\mathbb{Y}}

\DeclareMathOperator{\Arg}{Arg}

\def\AnswerYes{y}
\def\ShowLabelsVersion{n}         %%%%%% Version with defs. of refs, cites or final
\def\ShowChangesVersion{n}        %%%%%% Version with changes highlighted or final
\def\ShowAnnotationsVersion{n}    %%%%%% Version with annotations or final
\ifx\ShowLabelsVersion\AnswerYes
 \usepackage{showkeys}
\fi
\ifx\ShowAnnotationsVersion\AnswerYes
  \newcommand{\comment}[1]{{\color{blue}\textit{#1}}}
  \newcommand{\margin}[1]{\marginpar{\scriptsize\sffamily\bfseries{#1}}}
\else
  \newcommand{\comment}[1]{}
  \newcommand{\margin}[1]{}
  
\fi
\ifx\ShowChangesVersion\AnswerYes
  \usepackage{ulem}
  \newcommand{\delete}[1]{\sout{#1}}            % delete #1 (strike-out)
  \renewcommand{\emph}[1]{\textit{#1}}           % ulem overwrites def of \emph as
\else

  \newcommand{\sout}[1]{}
  \newcommand{\xout}[1]{}

  \newcommand{\delete}[1]{}
\fi
\begin{document}
\preprint{MITP/15-073}
\title{Description of light nuclei in pionless effective field theory using the stochastic variational method}

\author{Vadim Lensky}\email{lensky@itep.ru}
\affiliation{Institut f\"ur Kernphysik and PRISMA  Cluster of Excellence, Johannes Gutenberg Universit\"at Mainz,  D-55128 Mainz, Germany}
\affiliation{Theoretical Physics Division, School of Physics and Astronomy, University of Manchester, Manchester, M13 9PL, United Kingdom}
\affiliation{Institute for Theoretical and Experimental Physics, 117218 Moscow, Russia}
\affiliation{National Research Nuclear University MEPhI (Moscow Engineering Physics Institute), 115409 Moscow, Russia}

\author{Michael C.~Birse}\email{mike.birse@manchester.ac.uk}
\author{Niels R.~Walet}\email{niels.walet@manchester.ac.uk}
\affiliation{Theoretical Physics Division, School of Physics and Astronomy, University of Manchester, Manchester, M13 9PL, United Kingdom}

\date{\today}

\begin{abstract}
We construct a coordinate-space potential based on pionless 
effective field theory (EFT) with a Gaussian regulator. 
Charge-symmetry breaking is included through the Coulomb potential 
and through two- and three-body contact interactions. Starting with the effective field theory 
potential, we 
apply the stochastic variational method to determine the ground
states of nuclei with mass number $A\leq 4$. At 
next-to-next-to-leading order,
two out of three independent three-body parameters can be fitted to the three-body binding
energies. To fix the remaining one, we look for a simultaneous
description of the binding energy of $^4$He and the charge radii of 
$^3$He and $^4$He. We show that at the order considered we can find an acceptable solution,
within the uncertainty of the expansion. We find that the EFT expansion shows good agreement with empirical data within the estimated uncertainty, 
even for a system as dense as $^4$He.
\end{abstract}

\date{\today}
%\date{\parbox{\linewidth}{\centering%
%  \bigskip\bigskip
%  \today\endgraf%\bigskip
%\comment{REORDER THE REFERENCES BEFORE PUBLISHING}
%\bigskip
%  }}
\maketitle
%\thispagestyle{empty}
%\tableofcontents

\section{Introduction}
The idea of an effective field theory (EFT) is now firmly entrenched 
in physics: it provides a systematic way to describe a quantum system
using the relevant degrees of freedom at some low-energy scale,
combined with a systematic expansion in powers of that scale divided by a reference scale.
The use of such theories in nuclear physics takes its cue from the 
work of Weinberg \cite{Weinberg:1990rz,Weinberg:1991um}. The original
formulation is in terms of nucleons and pions and is based on 
expansions in ratios of small momenta or the pion mass to 
a typical QCD scale, such as the mass of the $\rho$ meson . 

For momentum scales that are even smaller than the pion mass $m_\pi$, 
as they are in the ground states of the lightest nuclei, 
we can reduce the degrees of freedom still further and work with a 
pionless EFT~\cite{Kaplan:1998xu,Chen:1999tn,Bedaque:2002mn}.
This is constructed in terms of contact interactions
involving two or more nucleons, ordered in powers of $Q=p/m_\pi$,
where $p$ is a typical nucleon momentum. At each order in the 
expansion new counter terms appear, and these need to be fixed 
from experimental data. A similar approach should also be applied
to other observables to derive effective operators that are 
consistent with the effective interactions.

A key feature of the nucleon-nucleon interaction is the strong
attraction in $S$-waves. This is signalled by the scattering lengths
which are unnaturally large compared to the scale of the underlying
physics, $1/m_\pi$. This means that the inverses of these scattering lengths
provide additional low-energy scales and the power counting must be
modified to take account of these scales. The leading $S$-wave contact terms 
are then of order $Q^{-1}$, which means that they can not be treated perturbatively, 
and have to be resummed to all 
orders. The resulting expansion of the scattering amplitude is just
the long-established effective range expansion \cite{Bethe}.

The first applications of this EFT to few-nucleon systems 
were to the triton \cite{Bedaque:1998kg}, and used the 
Skorniakov--Ter-Martirosian equation \cite{Skorniakov}. 
A disadvantage of this approach is that it does not provide easy 
access to the full wave function of the system. Calculations
of observables in the framework can become somewhat complex, as can be seen, 
for example, in the recent work of Vanasse \cite{Vanasse:2015fph}.

One alternative is to use an approach based on a Hamiltonian formulation of
the EFT. The pionless EFT has previously been treated by Platter \emph{et
al}.~\cite{Platter:2004zs} using the Yakubovsky equations,
and by Kirscher \emph{et al}.~\cite{Kirscher:2009aj}
using a variational approach based on the resonating-group
method~\cite{Tang:1978zz,Hoffmann:1987}. Such
variational methods  have a long tradition  in studies of few-nucleon systems 
with finite-range forces. (For a review, see 
Ref.~\cite{Kievsky:2008es}.) 

In our calculations, we employ one of these approaches, the stochastic variational method 
(SVM) \cite{Suzuki:1998bn}. This is a method for finding
bound-state solutions to the Schr\"odinger equation, which has proved
powerful in a variety of contexts. As in the case of the resonating-group method, this 
starts from a Hamiltonian formulation of the EFT, rather than a 
Lagrangian one. It is also easiest to apply when the short-range 
interactions are expressed in the form of a local potential in 
coordinate space. Kirscher \emph{et al.} achieved this by employing  a 
Gaussian regulator~\cite{Kirscher:2009aj}. 
We follow the same approach here since such
a regulator is especially well suited to the SVM. 

We work at next-to-next-to-leading order (NNLO) in the pionless EFT, with
a Hamiltonian that contains two- and three-nucleon interactions and
the Coulomb potential. Using the SVM to solve the Schr\"odinger 
equation provides bound-state wave functions as well as energies
for nuclei with $A=3$ and $4$. With these wave functions, we calculate
charge radii to the same order in the expansion. The basic calculations 
of the ground-state energies are similar to those of Kirscher 
\emph{et al.}~\cite{Kirscher:2009aj}; the main difference between 
our approaches is that we explore 
a consistent description of energies and other
observables at the same level of truncation. We also look more closely at 
ambiguities in the parameter fits.

Solving an EFT by use of the Schr\"odinger equation implicitly iterates 
all terms in the interaction to all orders. This is rather different 
from the strict perturbative expansion that has been implemented 
within the Lagrangian approach, e.g., \cite{Konig:2015aka}. Iterating
the potential in this way generates higher-order contributions to
observables that are in principle beyond the accuracy of our treatment. 
However, provided that the expansion of the EFT is converging, these 
contributions should be within the uncertainty of our truncation
of the theory. The iteration also means that the parameters are 
only implicitly renormalised by fitting them to observables. One has to be careful: such a 
procedure can break down if the regulator scale is taken above the 
scale of the underlying physics \cite{Epelbaum:2009sd,Birse:2009my}.

In our approach we first fit two-body potentials to low-energy nucleon-nucleon
phase shifts. The short-range potentials are regulated 
by a Gaussian function whose width plays the role of a cut-off scale.
Varying this width generates a set of potentials that all describe
the same two-body data. We use the dependence on the cut-off scale 
to determine
where our approach breaks down and to estimate the uncertainties on
our results. We also include charge-symmetry breaking (CSB), 
in particular from
the Coulomb potential, but also from the contact interaction
that is needed to renormalise the $pp$ scattering length 
\cite{Kong:1998sx,Kong:1999sf}.

We then add a three-body potential and examine the three-nucleon 
systems $^3$H and $^3$He. When Coulomb effects are included, 
a CSB three-body interaction is also
required~\cite{Vanasse:2014kxa}, leading to a potential with three
free parameters. Only two of these can be determined from the
ground-state energies. Varying the remaining parameter allows us to explore
 parametric relations between the energies 
and charge radii of nuclei with $A=3$ and $4$.

The article is structured as follows. First we set out, in
Sec.~\ref{sec:nopieft}, the details of the effective theory that we use,
and give expressions for the relevant two- and three-body potentials. 
We then outline, in Sec.~\ref{sec:svm}, the important features of 
the stochastic variational method (SVM), which we use to solve the 
Schr\"odinger equation for the bound states of three- and 
four-nucleon systems. The determination of the two-body parameters 
from scattering phase shifts is discussed in Sec.~\ref{sec:nn}. 
Then, in Sec.~\ref{sec:NNN}, we apply the SVM to the ground states of
${^3}$H, ${^3}$He and ${^4}$He, and we present our results for their
properties. We finally discuss some implications of our results in
Sec.~\ref{Sec:discussion}. 
Details of the Kohn variational principle used in the two-nucleon 
sector and an analysis of convergence are contained in the appendices.

\section{Pionless EFT}
\label{sec:nopieft}

We work with the pionless EFT at next-to-next-to-leading order (NNLO), using 
the power counting for systems with anomalously large $S$-wave 
scattering lengths 
\cite{Kaplan:1998xu,Bedaque:2002mn}. To NNLO, the effective Lagrangian
consists of two- and three-body contact interactions containing 
up to two derivatives. The resulting nucleon-nucleon interaction 
can be expressed as a momentum-space potential with the following 
form (see, for example, Refs.~\cite{Ordonez:1995rz,Kirscher:2009aj}):
\begin{equation}
\begin{split}
V_{NN}=
&\, \frac{1}{2}\sum_{i\neq j}\Bigl[
C_1 + C_2\,\svec \sigma_i \cdot\svec\sigma_j + D_1 q^2 + D_2 k^2 + \svec\sigma_i\cdot\svec\sigma_j\left(D_3 q^2+D_4 k^2\right)\\
&+\frac{1}{2}D_5\left(\svec\sigma_i+\svec\sigma_j\right)\cdot \svec q\times \svec k + D_6 (\svec q\cdot \svec\sigma_i)( \svec q\cdot \svec \sigma_j)
+ D_7 (\svec k\cdot \svec \sigma_i)( \svec k\cdot \svec \sigma_j)
\Bigr]\,,
\label{eq:vnn1}
\end{split}
\end{equation}
where for each pair $ij$, $\svec q=\svec p_i - \svec p_i'$ 
and $\svec k=(\svec p_i+\svec p_i')/2$ are defined in terms of
the initial and final centre-of-mass momenta of one of the nucleons,
$\svec p_i$ and $\svec p_i'$.
Note that $\svec q$ is just the difference between the relative momenta 
of the two nucleons before and after the interaction,
whereas in the two-body centre-of-mass frame $\svec k$ is the average 
of these momenta. 

For systems with very strong $S$-wave scattering at low energies, the large
scattering lengths should be counted as of order $Q^{-1}$ 
\cite{Kaplan:1998xu,Bedaque:2002mn}. That is the case for nucleon-nucleon scattering, and thus the leading-order (LO) $S$-wave contact interactions 
($C_1$ and $C_2$) are of order $Q^{-1}$ and hence need 
to be treated nonperturbatively. The higher-order interactions 
are also enhanced relative to a na\"ive power counting, and their orders can be obtained from an 
expansion around a non-trivial fixed point of the renormalisation 
group \cite{Birse:1998dk}. In particular, the subleading $S$-wave 
contact interactions are promoted by two powers of $Q$ relative to na\"ive
dimensional analysis, and terms that couple $S$-waves to other 
channels, such as the $S$--$D$ mixing, are promoted by one power. 

At next-to-leading order (NLO), we find the momentum-dependent $S$-wave interactions, $D_1$ to $D_4$, which
enter at $\mathcal{O}(Q^0)$. At NNLO or $\mathcal{O}(Q^1)$, 
the tensor interactions $D_6$ and $D_7$ contribute to $S$--$D$ mixing. The $D_5$ term describes the 
spin-orbit interaction, and so does not contribute in $S$-waves interactions, hence we ignore it as 
it enters only at $\mathcal{O}(Q^2)$. We also omit the $D_7$ term, which generates a non-local 
tensor interaction, since its effects cannot be distinguished from 
those of the local $D_6$ term to the order we work here; disentangling them would require data on $P$ waves, whose amplitudes start at $\mathcal{O}(Q^2)$. 
 
As a further simplification, we also make use of the fact that the total
spin and isospin of the nucleon-nucleon partial waves are correlated
and so there is some freedom in the choice of the spin 
and isospin operators, as pointed out in, for example,
Refs.~\cite{Weinberg:1990rz,Ordonez:1995rz}. A recent discussion can be
found in Ref.~\cite{Gezerlis:2014zia}, which makes a slightly different
choice from ours. Since the tensor interaction couples $S$ and $D$ 
waves with spin-1 and isospin-0, we choose an isospin structure that
projects the $D_6$ term onto isospin-0. Finally, for the central
interactions, we keep only terms with the isospin-dependent form 
$\svec\tau_i\cdot\svec\tau_j$. 
These choices simplify the fitting of the potential to empirical data.

In order to to make use of this potential in the SVM, which works 
in coordinate space, we take its 
Fourier transform. At this step we introduce a local Gaussian regulator,
\begin{equation}
G(r,\sigma)=\exp\left(-\frac{1}{2}\frac{r^2}{\sigma^2}\right),
\end{equation}
similar to that used by Kirscher \textit{et al.}~\cite{Kirscher:2009aj}.
The resulting coordinate-space potential has the form
\begin{equation}
\begin{split}
V_{NN}=& \sum_{i<j}\Bigl\{
G(r_{ij},\sigma)\left(A_1+A_2 \svec{\tau_i}\cdot\svec{\tau_j} \right)
+r_{ij}^2G(r_{ij},\sigma)\left(A_3+A_4 \svec{\tau_i}\cdot\svec{\tau_j} \right)\\
&
\qquad+\left\{\nabla_{ij}^2,G(r_{ij},\sigma)\right\}
\left(A_5+A_6 \svec{\tau_i}\cdot\svec{\tau_j} \right)
\\
&\qquad+G(r_{ij},\sigma)A_7\left(1-\svec{\tau_i}\cdot\svec{\tau_j}\right)
\left[3(\hat{\svec r}_{ij}\cdot\svec\sigma_i)(\hat{\svec r}_{ij}\cdot\svec\sigma_j)-(\svec\sigma_i\cdot\svec\sigma_j)\right]\Bigr\}\,,
\end{split}
\label{eq:vnn}
\end{equation}
where $\svec r_{ij}=\svec r_i - \svec r_j$, $\hat{\svec r}_{ij}$ is the
corresponding unit vector, and $\{\nabla^2,G(r,\sigma)\}$ stands for the commutator of the two operators acting on the wave function. The parameters $A_i$ are linear combinations 
of the $C_i$ and $D_i$, showing the mixing between orders in a potential model. 
Note that the tensor interaction has been explicitly projected out of the isospin-1 channel. The non-local terms give rise to derivative operators in Eq.~\eqref{eq:vnn}, see, for example, Ref.~\cite{Nagels:1977ze}.

We also include the long-range Coulomb potential that acts between protons,
\begin{equation}
V_{pp}^\mathrm{C}=\frac{\alpha_\mathrm{em}}{r_{ij}}\,,
\end{equation}
where $\alpha_\mathrm{em}$ is the fine structure constant. 
Two choices of power counting are possible for this potential.
For the first choice one identifies the scale related to its strength, 
$\alpha_\mathrm{em}M$, with $M$ being the nucleon mass,
as a low-energy scale, of order $Q$. 
This makes the Coulomb potential $\mathcal{O}(Q^{-1})$
and so it should be resummed to all orders, as required in order to
properly describe, for instance, very-low-energy proton-proton
scattering~\cite{Kong:1998sx,Kong:1999sf}. In this scheme, there
is a $pp$ CSB contact interaction proportional to $\alpha_\mathrm{em}$ 
which also appears at order $\mathcal{O}(Q^{-1})$ and is needed
to renormalise the $pp$ scattering length. This approach was
used in Ref.~\cite{Ando:2010wq} and more recently by Vanasse 
\cite{Vanasse:2014kxa} in studies of three-nucleon systems. 
Applications to these systems require an additional CSB three-body
force of order $\mathcal{O}(Q^0)$.

However the typical momenta in the three-body bound states are
significantly larger than $\alpha_\mathrm{em}M$, suggesting a second choice where 
the Coulomb interaction can be treated perturbatively, as proposed by
Rupak and Kong~\cite{Rupak:2001ci} and confirmed in more detailed work
by K\"onig \textit{et al.}~\cite{Konig:2015aka}. This can be 
implemented in the EFT framework by taking $\alpha_\mathrm{em}M$ 
to be $\mathcal{O}(Q^{2})$, making the Coulomb potential itself of 
order $Q^0$. In this counting scheme the CSB two- and 
three-body contact interactions appear at one order higher in $Q$ than
in the Kong and Ravndal version: 
$\mathcal{O}(Q^0)$ and $\mathcal{O}(Q^{1})$ respectively.

In either case, we need to take account of the CSB 
contact interactions. In the two-nucleon sector, we add the following
interactions to the $pp$ channel:
\begin{equation}
V_{pp}^\mathrm{CSB}=G(r_{ij},\sigma)A_1^\mathrm{CSB}+r_{ij}^2G(r_{ij},\sigma)A_3^\mathrm{CSB}+\left\{\nabla_{ij}^2,G(r_{ij},\sigma)\right\}A_5^\mathrm{CSB}\,.
\label{eq:vnniv}
\end{equation}
These forces modify the strong interaction between two protons and
renormalise the effects of the Coulomb potential. Since we  
implicitly treat that potential to all orders, we include 
a CSB contribution to the $pp$ effective range and we find that it 
is needed to give a good description of the low-energy $pp$ phase shift.
However, in the counting scheme where the Coulomb potential is 
treated perturbatively, that term would be of order $Q^2$ and so, strictly
speaking, it would be beyond the accuracy of our current treatment.

We solve the Schr\"odinger equation with the regulated potential and fit the coefficients $A_i$ and $A_i^\mathrm{CSB}$ 
to data from
low-energy nucleon-nucleon scattering. Since the potential is treated to all orders in this procedure, 
we rely on an implicit renormalisation of the parameters, as outlined by Gasser and Leutwyler 
\cite{Gasser:1982ap} and discussed further in Ref.~\cite{Epelbaum:2009sd}. Resumming the potential 
generates terms in the amplitude that are beyond the order of our truncation. It has long been known that 
these can cause problems if the regulator has too short a range \cite{Scaldeferri:1996nx,Phillips:1996ae}.
However, provided the momentum scale of the regulator is not taken above the breakdown scale of the EFT,
these higher order terms are within the error of the truncation. On the other hand regulating the theory at much lower scales or longer distances 
can generate large artefacts of the cut-off. Reconciling these conflicting
demands suggests that the regulator scale should be chosen just below the
breakdown scale of the EFT \cite{Epelbaum:2009sd,Birse:2009my}.

In the case of the pionless EFT, we expect the appropriate scale to 
be of the order of $1/m_\pi$. The precise value will depend on
the the chosen form of the regulator. For instance, a lower bound 
of $R=1.3$~fm was found for a sharp radial cut-off in Ref.~\cite{Scaldeferri:1996nx}. 
Here we explore several choices for the parameter $\sigma$ in the 
range 0.6 to 1.2~fm. We examine how the fitted values of the 
coefficients vary with the width of the Gaussian regulator, $\sigma$, and exclude the region where these coefficients
start to become ``unnaturally'' large. This is contrast to
the recent work by Kirscher and Gazit \cite{Kirscher:2015zoa}
who use a similar potential to ours but with cut-offs corresponding
to $\sigma$ in the range 0.08 to 0.25~fm. Treating the NLO potential 
to all orders with such cut-offs requires a high degree of fine-tuning,
and runs the risk of generating unphysical features such as producing 
an additional, deeply bound, three-nucleon state.

We now turn our attention to three-nucleon systems. In these,
three-body forces have been shown to be needed at leading order 
in the pionless EFT in order to renormalise observables such as 
the triton binding energy~\cite{Bedaque:1998kg}.
This, in turn, can be related to the Efimov effect --- a geometric
sequence of bound states seen in the symmetric $S$-wave
channels of three particles interacting via contact
interactions~\cite{Efimov:1970zz}. In EFT treatments of such
systems, the leading three-nucleon interaction exhibits
limit-cycle behaviour \cite{Bedaque:1998kg} and so  has 
an anomalous power counting of $\mathcal{O}(Q^{-1})$ 
\cite{Barford:2004fz}.

At LO there is only one independent spin-isospin structure 
in the three-body contact interaction since all possible 
combinations of spin and isospin operators give equivalent 
results when they act on antisymmetrised wave functions
\cite{Bedaque:2002yg,Epelbaum:2002vt}. Taking advantage of
this equivalence, we work with the isospin- and spin-independent 
expression,
\begin{equation}
V_{3N}^\mathrm{LO}=\sum_{i<j<k}E_1\,,
\end{equation}
for the LO three-body potential. We fix the strength of this 
interaction, $E_1$, by fitting it to the triton binding energy.

At NLO, a $\mathcal{O}(Q^0)$ three-nucleon force is needed only 
in order to properly renormalise the effects
of the two-body effective range; since this does not introduce 
either new spin-isospin structures or dependence on the nucleon 
momenta, it cannot be separated from the LO potential. 

Finally, at NNLO we need to include three-nucleon forces that are
quadratic in the nucleon momenta. The corresponding
potential contains ten independent structures~\cite{Girlanda:2011fh}:
\begin{equation}
\begin{split}
V_{3N}^\mathrm{NNLO} = &
\sum_{i\neq j\neq k}\Bigl[
\ q_i^2\left(F_1+F_2\svec\tau_i\cdot\svec\tau_j 
+ F_3\svec{\sigma}_i\cdot\svec{\sigma}_j
+F_4\svec{\sigma}_i\cdot\svec{\sigma}_j\svec\tau_i\cdot\svec\tau_j
\right)\\
&\qquad
+\left[3(\svec q_i\cdot\svec\sigma_i)(\svec q_i\cdot\svec\sigma_j)
-q_i^2\right](F_5+F_6\svec\tau_i\cdot\svec\tau_j)\\
&\qquad + \frac{1}{2}(\svec\sigma_i+\svec\sigma_j)\cdot 
\svec q_i\times(\svec k_i - \svec k_j)
(F_7+F_8\svec\tau_j\cdot\svec\tau_k)\\
&\qquad
+(\svec k_i\cdot\svec\sigma_i)(\svec k_j\cdot\svec\sigma_j)(F_9+F_{10}\svec\tau_i\cdot\svec\tau_j)\Bigr]\,,
\end{split}
\end{equation}
where $\svec q_i=\svec p_i - \svec p_i'$ and 
$\svec k_i = (\svec p_i + \svec p_i')/2$ are defined in terms of
the centre-of-mass momenta of one nucleon, as in the two-body case.
However, some of these structures couple only to higher partial waves 
which do not receive the enhancement present in the spatially symmetric 
$S$-wave channel~\cite{Griesshammer:2005ga}. They therefore contribute 
only at higher orders and 
we neglect them. These are the terms proportional to $F_{5,6}$ and 
$F_{9,10}$, the three-body tensor interaction and its recoil 
corrections, and the ones containing $F_{7,8}$, the three-body 
spin-orbit force. 

This leaves four momentum-dependent structures proportional to 
$F_{1\dots 4}$. These are equivalent if the nuclear wave function 
is symmetric under exchange of the momenta of any two nucleons.
Since the ground-state wave functions of the
nuclei we study --- ${^3}$H, ${^3}$He, ${^4}$He --- are 
dominated by the spatially symmetric $S$-wave channel,
we can choose one of the terms $F_{1\dots 4}$ to represent the effect of the 
$\mathcal{O}(Q^1)$ three-nucleon potential.
Guided by this, we choose the following symmetric form,
\begin{equation}
V_{3N}^\mathrm{NNLO}=\sum_{i<j<k}F_1 (q_i^2+q_j^2+q_k^2)\,.
\end{equation}

As in the two-body case, we express this potential in 
coordinate space using a Gaussian regulator, 
with the additional requirement that it 
should depend on relative momenta only and be symmetric with 
respect to interchange of any two particles. 
The use of a local form for the regulator can break the equivalence
of the different possible spin-isospin structures, 
as observed by Lynn \textit{et al.}~\cite{Lynn:2015jua}.
More specifically, as those authors point out, this dependence on
the spin-isospin structure shows up in the $P$-wave states, where
three-body forces are of higher-order than considered here. 

For simplicity, we choose the range parameter of our regulator 
to be the same as in the two-body 
potential. This leads to the following coordinate-space form for
the three-body potential, including both LO and NNLO terms:
\begin{equation}
V_{3N}=\sum_{i<j<k}
\left[B_1 + B_2\frac{1}{\sigma^2}\left( r_{ij}^2+r_{ik}^2+r_{jk}^2\right)\right]\exp\left(-\frac{1}{2\sigma^2} \left( r_{ij}^2+r_{ik}^2+r_{jk}^2 \right)\right)\,,
\label{eq:vnnn}
\end{equation}
This can also be expressed in Jacobi coordinates,
\begin{equation}
\svec\xi_1=\frac{1}{\sqrt{2}}(\svec r_i - \svec r_j),\quad \svec \xi_2=\sqrt{\frac{2}{3}}\left(\frac{1}{2}(\svec r_i+\svec r_j)-\svec r_k\right)\,;
\end{equation}
with the help of the relation
\begin{equation}
\xi_1^2+\xi_2^2 =\left( r_{ij}^2+r_{ik}^2+r_{jk}^2\right)/3\,.
\end{equation}

Finally, the inclusion of the Coulomb potential in the proton-proton 
sector leads to CSB in the three-nucleon interaction.
In particular, renormalising the three-body interaction in the presence
of a Coulomb force leads to a contact term proportional to 
$\alpha_\mathrm{em}$.
As discussed above, iterating the Coulomb potential corresponds
to treating $\alpha_\mathrm{em}M$ as being of order $Q$, and so this
interaction appears at NLO or $\mathcal{O}(Q^0)$, as in the
work of Vanasse~\cite{Vanasse:2014kxa}. Alternatively, in a purely
perturbative treatment like that of K\"onig 
\textit{et al.}~\cite{Konig:2015aka}, $\alpha_\mathrm{em}M$ would be
assigned an order $Q^2$, which would demote this term to NNLO
or $\mathcal{O}(Q^1)$. 

Again, in either scheme, this CSB interaction should be included 
in our Hamiltonian. It has the same form as the leading three-nucleon
interaction but acts only if any two of the three nucleons involved
are protons:
\begin{equation}
V_{ppn}^\mathrm{CSB}=B_\mathrm{CSB}\exp\left(-\frac{3}{2}\frac{\xi_1^2+\xi_2^2}{\sigma^2}\right)\,.
\label{eq:vnnncsb}
\end{equation}

We thus have, in addition to the seven parameters from the two-nucleon
sector, three three-nucleon strengths. Determining these uniquely
would require not just three-body ground states but also scattering
observables, and these are not accessible to our current variational
method. We therefore choose to fit two of the three-body parameters 
to the ground-state energies of ${^3}$H and ${^3}$He. 
For example, for any value of $B_1$, we can fix $B_2$ from
the triton binding energy and $B_\mathrm{CSB}$ from the
binding energy of ${^3}$He. This leads to parametric 
relationships between 
the quantities involved in the calculation, for example
the values of $B_1$ and $B_2$, and between the results 
for observables such as charge radii and the binding energy of ${^4}$He.
Going further and fixing the remaining three-body parameter 
from three-nucleon strong-interaction observables will require an
extension of our method to describe scattering states.

\section{Stochastic variational method}
\label{sec:svm}
We use the stochastic variational method (SVM)~\cite{Suzuki:1998bn} 
to solve the Schr\"odinger equation for few-nucleon systems. 
It is one of the very accurate methods that can be brought to bear 
on such problems, and has a proven track record. A particularly 
useful feature is that its trial functions can be expressed in a 
Gaussian basis, which greatly simplifies the calculations when working with 
a Gaussian potential. The method is based on a stochastic
algorithm that determines the parameters of a set of trial wave functions.
These wave functions form a (non-orthogonal) basis in which the
Hamiltonian is diagonalised to give the ground-state energy. 

We set up our basis of trial functions for a system of $N$ 
nucleons such that they all 
have the appropriate total angular momentum, total spin 
(in the absence of spin-orbit forces), and total momentum.
For the work described here we choose the form
\beq
\ket{\psi_{JJ_zLS}^\alpha}=\sum\limits_{M,S_z}
\langle LM\,SS_z|JJ_z\rangle\,
\ket{f_{LM}}\ket{SS_z,\alpha}\,,
\label{eq:svm_wavefn}
\eeq
where the orbital angular momentum $L$ and the spin $S$ are coupled 
to a total $J$ with  $z$-projection $J_z$.
Here, the index $\alpha$ reflects the fact that there
are multiple different ways 
to couple $N$ spin-$\frac{1}{2}$ states to a total 
spin of $S$. The parity of the trial functions is
fixed by the choice of appropriate orbital angular momenta $L$.

We choose to work in a proton-neutron basis as opposed to an isospin
basis. This means that we treat protons and neutrons as distinct
species of particles, characterised only by coordinate and spin
degrees of freedom. The inclusion of the Coulomb and CSB forces is 
thus straightforward. This choice also saves some calculational effort
as the few-nucleon wave functions need to be antisymmetrised only 
over protons and neutrons separately. On the other hand, this
calculational gain is offset by the fact that the isospin-dependent
potentials in Eq.~\eqref{eq:vnn} need to be expressed in the form
of proton-neutron exchange potentials.

The spin part of the wave function is constructed in two steps: 
first, the allowed spins of $N_p$ protons and $N_n=N-N_p$ neutrons
are coupled to total proton and neutron spins, $S_p$ and $S_n$. 
These individual spin states have the form ($\mu=p,n$)
\beq
\ket{S_\mu S_{\mu z},\,N_\mu,\,\alpha_\mu}=
\sum\limits_{\left\{\lambda\right\}}D^{S_\mu S_{\rho z}(N_\mu,\,\alpha_\mu)}_{\left\{\lambda\right\}}
\bigotimes_{i=1}^{N_\mu} \ket{\lambda_i}\,.
\label{eq:spinwf}
\eeq
Here the sum runs over all sets 
$\left\{\lambda\right\}=\{\lambda_1,\ldots,\lambda_{N_\mu}\}$ 
of single-particle spin projections $\lambda_i$ such that
$\sum\lambda_i=S_{\mu z}$, and $\alpha_{\mu}$ enumerates the different
coupling schemes. The coefficients $D^{SS_z(N,\ \alpha)}$ can be
calculated either by using Clebsch-Gordan coefficients to add spins one
at a time,  or by diagonalising the operator $S^2$ in the space of all
possible sets $\{\lambda\}$.
In the second step the functions $\ket{S_pS_{pz},\,N_p,\,\alpha_p}$ and $\ket{S_nS_{nz},\ N_n,\ \alpha_n}$ are  coupled to the 
required total spin $S$ and projection $S_z$,
\beq
\ket{SS_z,\,\alpha}=\sum\limits_{S_{pz}\,S_{nz}}
\langle S_pS_{pz}\,S_nS_{nz}|SS_{z}\rangle\,
\ket{S_pS_{pz},\,N_p,\,\alpha_p}\ket{S_nS_{nz},\ N_n,\ \alpha_n}\,.
\label{eq:spinwf_coupled}
\eeq

To allow for easy evaluation of overlap integrals
and potential matrix elements, we choose a generalised Gaussian 
for the coordinate part of each trial function. This has the form 
\beq
\brkt{\{\svec{x}\}}{f_{LM}}_{\mat A, u, K}=f_{LM}(\{\svec x\},\mat A, u, K)=v^{2K}\Y_{LM}(\svec v)\exp\left(-\frac{1}{2}\mat A^{ij} {\svect x_i} \svec x_j\right),
\label{eq:G_trial_fn}
\eeq
where $\{\svec x\}=\{\svec x_i,\,i=1,\dots, N-1\}$ are the Jacobi
coordinates describing the relative motion, and
$\mat A$ is a symmetric positive-definite $(N-1)\times (N-1)$ matrix.
Here the $\Y_{LM}$ are the solid spherical harmonics,
 $\Y_{LM}(\svec v)=v^L Y_{LM}(\svec{\hat v})$, with
 $\svec v=\sum_i u^i \svec x_i$.
The ``direction vector'' $u=\left(u^i,\ i = 1,\dots, N-1\right)$ 
provides a way to add angular dependence to the variational wave
functions and leads to a rather simple form for ones that have
non-zero orbital angular momentum. The factor $v^{2K}$ with
$K$ being a non-negative integer is an additional means to
adjust the shape of the variational wave functions. Variational parameters are
provided by the elements of the matrix $\mat A$, the vector $u$, and the integer
$K$.

The trial functions thus constructed are then antisymmetrised,
as required by the Pauli exclusion principle. The antisymmetrisation
operator acts on $\ket{\psi_{JJ_zLS}^\alpha}$ to generate
a sum of terms analogous to \eqref{eq:spinwf}-\eqref{eq:G_trial_fn},
but with $\mat A$ and $u$ replaced by $\mat P^\mathsf{T} \mat A\mat P$
and $\mat P^\mathsf{T}u$, respectively,
where $\mat P=\mat J \mat C_\mathcal{P} \mat J^{-1}$ with $\mat J$
being the transformation from single-particle to Jacobi coordinates
and $\mat C_\mathcal{P}$ the permutation matrix corresponding
to a particular permutation $\mathcal{P}$ of the coordinates.
Note that elements corresponding to the transformation of the
centre-of-mass coordinate are omitted from $\mat P$ which
thus is an $(N-1)\times (N-1)$ matrix.
The spin indices $\{\lambda_i\}$ have to be permuted too;
note, however, that they are spin {\it labels}
as opposed to the spin {\it coordinates} and therefore they
are transformed under the inverse of the permutation $\mathcal{P}$.

With these choices both the basis functions and potentials have 
a Gaussian (or Gaussian times a power) dependence on the relative 
coordinates. This allows for a very efficient calculation of the 
matrix elements. All of the integrals over Gaussians can be reduced 
to algebraic calculations involving the inverse of the 
matrix $\mat A$. Since the latter is a positive-definite matrix,
it can also be inverted very efficiently using, for example, 
the Cholesky decomposition. Angular dependence as in
Eq.~\eqref{eq:G_trial_fn} can also be taken into 
account analytically (see Ref.~\cite{Suzuki:1998bn} for details).

The individual trial functions (\ref{eq:svm_wavefn}) are labelled 
by the variational parameters $\{\alpha,\mat A, u, K\}$, 
values for which are generated by the stochastic process. 
In general, a single wave function of this type is 
too simple to give a good approximation to the ground state, but
they can be used as basis functions for a much richer
variational Ansatz. The wave function for the ground state is taken 
to be a linear combination of these basis functions (note we select
the stretched state, $J=J_z$),
\beq
\ket{\Psi_0}=\sum\limits_{i}c_i\ket{\psi_i}=\sum\limits_{i}c_i\ket{\psi_{JJL_iS_i}^{\alpha_i}(\mat A_i, u_i, K_i)}.
\eeq
Varying the coefficients $c_i$ is equivalent to solving
the generalised eigenvalue problem
\beq
\mat H^{ij}c_j=E\mat N^{ij}c_j,\quad i,j = 1,\dots,m\,,
\label{eq:gevp}
\eeq
where $\mat H$ and $\mat N$ are the Hamiltonian and  
overlap matrices in a given basis,
\beq
\mat H^{ij}=\bra{\psi_i}H\ket{\psi_j},\quad 
\mat N^{ij}=\brkt{\psi_i}{\psi_j}\,.
\eeq
Note that two of these basis functions with different parameters are 
not, in general, orthogonal, and so $\mat N$ is not a diagonal matrix.
This can lead to problems of approximately over-complete basis sets, 
as discussed below.

As the number of random parameters grows with the size of the basis 
(and also with the number of particles), it is as a rule too costly 
to vary the parameters of more than one basis 
state at a time. In the simplest implementation, we therefore assume
that we have determined $m$ good basis states, and try to add one 
additional function of the same form but with different parameters. 
We therefore know the lowest eigenvalue $E_0$ of the generalised
eigenvalue problem Eq.~(\ref{eq:gevp}) for the current basis set.
When a single state is added to the basis, there is a very
efficient method for solving the resulting eigenvalue problem with $m+1$ 
basis states~\cite{Suzuki:1998bn}. This allows us to try several 
states, and keep the one that lowers the energy $E_0$ by the largest
amount. 

This approach can still lead to very large basis sets, where many states
contribute very little to the overall binding. Another useful 
strategy is thus basis refinement: a stochastic process that removes 
basis states with a probability inversely proportional to their
contribution in the wave function under study.

Even with basis refinement, we can find states that are quite similar in our basis.
This can lead to numerical difficulties. States that are rather similar can 
nonetheless make important contributions to the wave function through
their differences, but in the worst case such states
lead to very small eigenvalues  of the overlap matrix $\mat N$, 
which in turn lead to instability in calculation of the energy eigenvalues.
An efficient way around this is to use the singular-value 
decomposition of $\mat N$ instead of attempting to find 
its inverse, and to only calculate the inverse in the subspace of singular values
that are sufficiently different from zero.

Apart from this rather common effect from non-orthogonal basis states,
a further problem can arise from the antisymmetrisation of
our basis states. If a particular trial state is
nearly Pauli-blocked, the norm of such a
state, after antisymmetrisation, is extremely small.
Due to the rounding errors present in any numerical calculation,
this can result in negative eigenvalues of the overlap matrix $\mat N$,
making the calculation unstable --- anyway it leads to a high condition number
for the matrix $\mat N$,
which should be avoided. Therefore we reject such
states whenever our stochastic principle suggests them.

\section{Nucleon-nucleon potential}
\label{sec:nn}

In the $pn$ sector, there are five physical observables we 
should be able to describe working to NNLO. These are: 
the singlet and triplet $pn$ scattering lengths $a_{s,t}$, 
the corresponding effective ranges $r_{s,t}$, and the strength of
the triplet $S-D$ mixing. There is some freedom of choice for the
last of these; here we take the value of the mixing angle $\epsilon_1$ 
at $T_\mathrm{lab}=10$~MeV (corresponding to a relative momentum of
$k=0.347~\mathrm{fm}^{-1}$). 

On the other hand, the potential in Eq.~(\ref{eq:vnn}) has seven
parameters, which means that, in order to fit it to these five
quantities, we need to impose two constraints. Our local potential 
can generate scattering in the $P$-waves 
even though this is of higher than NNLO in our counting. 
Appropriate constraints should therefore require the $P$-wave phase 
shifts to be small. Similar constraints were imposed in
Ref.~\cite{Kirscher:2009aj}. Taking advantage of the fact the $P$-wave
scattering is weak at low energies, it is convenient to demand that 
the $^1P$ and $^3P$ scattering amplitudes 
in the Born approximation vanish at some small finite momentum,
which we take to be $k=0.4~\mathrm{fm}^{-1}$. Since the $P$-wave 
phase shifts vanish as $k^2$ as $k\rightarrow 0$, these constraints 
ensure that they remain very small over the whole low-momentum region.
(Note that since we take the tensor potential to be isoscalar, 
the scattering produced by our potential is the same in all the triplet
$P$-waves, ${^3}P_{0,1,2}$.)
This leaves us with two parameters to fit to the spin-singlet 
(${^1}S_0$) channel, and three to the spin-triplet 
(${^3}S_1-{^3}D_1$) channels. 

With the Coulomb and CSB
strong forces acting in the $pp$ channel, we also want
to be able to reproduce two additional observables,
namely, the $pp$ scattering
length $a_{s}^{pp}$ and effective range $r_{s}^{pp}$.
Our CSB potential, Eq.~\eqref{eq:vnniv}, has three additional
parameters; to eliminate one of them, we demand that the $^3P$ scattering
amplitude (which is again the same in all the triplet $P$-waves)
in the Born approximation vanishes at $k=0.4~\mathrm{fm}^{-1}$
in the $pp$ channel as well. This leaves us with two CSB parameters 
to fit to the $pp$ channel.

\begin{figure}[th]
\includegraphics[width=\textwidth]{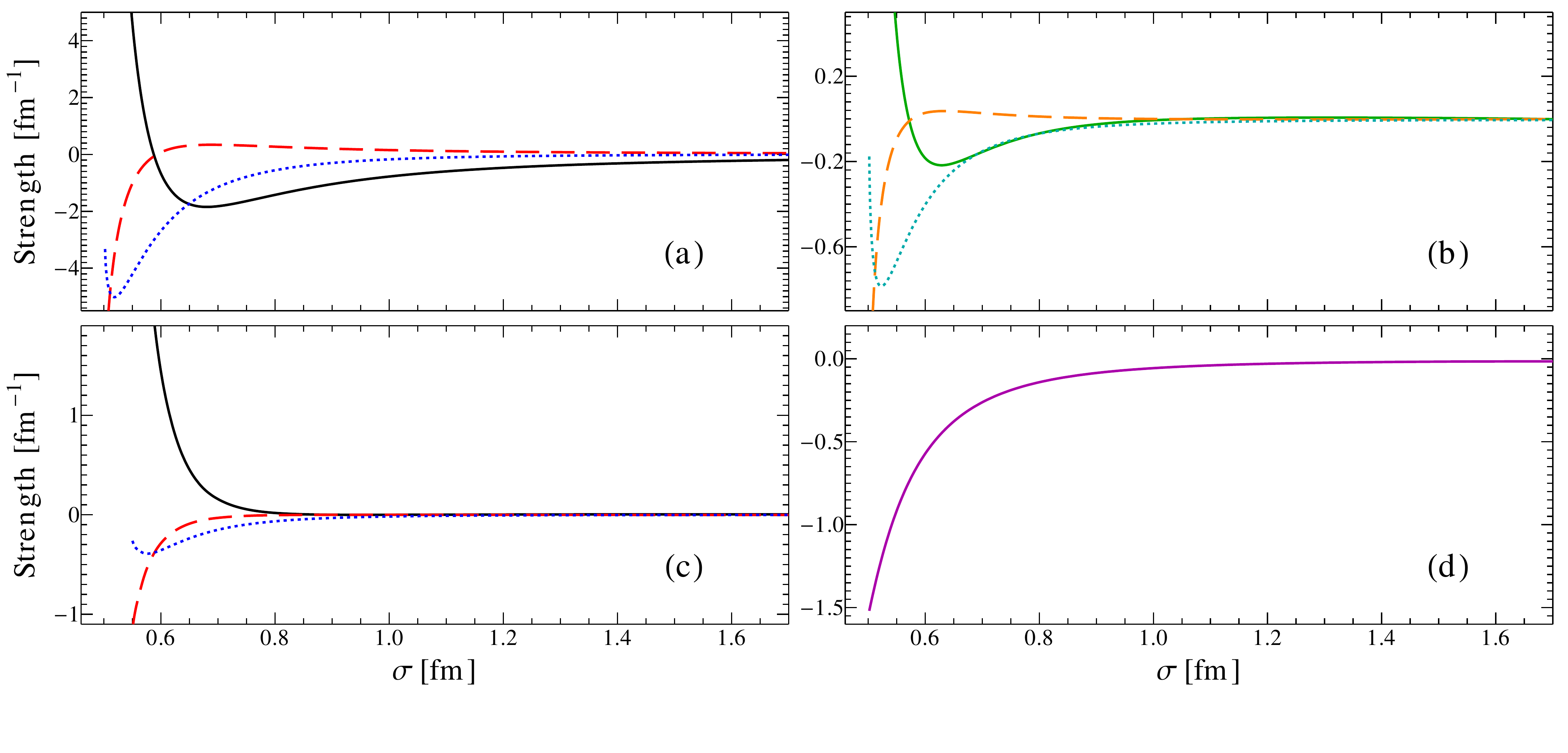}
\caption{(Color online) Running of the parameters of the $NN$ potential 
as a function of the range $\sigma$.
(a) $A_1$ (black solid curve), $A_3\sigma^2$ (red dashed curve),
$A_5/\sigma^2$ (blue dotted curve).
(b) $A_2$ (green solid curve), $A_4\sigma^2$ (orange dashed curve),
$A_6/\sigma^2$ (cyan dotted curve).
(c) $A_1^\mathrm{CSB}$ (black solid curve), $A_3^\mathrm{CSB}\sigma^2$
(red dashed curve), $A_5^\mathrm{CSB}/\sigma^2$ (blue dotted curve).
(d) $A_7$ (solid magenta curve). All parameters have been multiplied 
with appropriate powers of $\sigma$ so that they have units of $\mathrm{fm}^{-1}$.
}
\label{fig:potentials_running}
\end{figure}

We determine the parameters of the potential through the following 
fitting strategy. 
We solve the two-body Schr\"odinger equation to get the 
scattering amplitude using a complex version of the Kohn 
variational method \cite{Miller1987,McCurdy1987,Schneider:1988zz},
as outlined in Appendix~\ref{sec:Kohn}.
Starting from the smoothest cut-off (taken to be $\sigma=1.7$~fm),
we adjust the parameters of the potential until the resulting
values for the scattering observables match their empirical
values~\cite{Wiringa:1994wb,Stoks:1993tb,nn-online},
\begin{equation}
a^{pn}_{{^1}S_0}=-23.75~\mathrm{fm},\  a^{pn}_{{^3}S_1}=5.42~\mathrm{fm},\
r^{pn}_{{^1}S_0}=2.81~\mathrm{fm},\ r^{pn}_{{^3}S_1}=1.76~\mathrm{fm},\
\epsilon_1(10\;\mathrm{MeV})=1.1592^\circ\,
\end{equation}
and
\begin{equation}
a^{pp}_{{^1}S_0}=-7.806~\mathrm{fm},\ r^{pp}_{{^1}S_0}=2.79~\mathrm{fm}\,
\end{equation}
in the $pn$ and the $pp$ channels, respectively.
We then decrease the value of $\sigma$ in steps, using the previously
determined potential parameters as the starting point for each new search.
The results of this procedure are shown in
Fig.~\ref{fig:potentials_running}. From this, one can see that the
parameters depend relatively weakly on the cut-off $\sigma$ for
the longer-range  potentials, with $\sigma \gtrsim 0.8$~fm.

In contrast, for shorter-range potentials, the parameters become very 
strongly dependent on $\sigma$ and they grow rapidly to unnaturally 
large values for $\sigma\lesssim 0.6$~fm, with a slightly higher bound for the CSB interaction.
For these cutoffs, there have 
to be very large cancellations in the scattering amplitude between the
different terms in the potential. As a result there is a high degree of 
fine-tuning needed in order to reproduce the observables.
(This is also the region in which the calculation becomes numerically
more difficult, with the number of basis functions needed in the Kohn
method for a converged solution growing from 12 to over 20.)

This unstable growth of the couplings signals the expected breakdown 
of this treatment when the cut-off is taken beyond the domain of 
validity of the EFT. For our Gaussian regulator, we can estimate 
the breakdown scale to be of the order of $\sigma\simeq 0.6$~fm. 
Noting that the Gaussian cut-off has an effective range of
about $2$--$3\sigma$, this seems consistent with the 
scale of the underlying physics being $1/m_\pi$ and with the 
results of  Scaldeferri \emph{et al.}~\cite{Scaldeferri:1996nx}, 
who found a breakdown at $R\simeq 1.3$~fm for a sharp cut-off. 

We therefore work with four values of $\sigma$ in our few-body
calculations, namely, $\sigma=0.6,\ 0.8,\ 1.0$, and $1.2$~fm. 
The three larger values allow us to examine the sensitivity of our 
results to the choice of regulator and hence to estimate uncertainties 
on our results. The lowest value is there to check what happens at 
the breakdown point of our approach. The resulting values for 
the parameters $A_i$ and $A_i^\mathrm{CSB}$ from these fits are given in
Table~\ref{tab:potentials} and Table~\ref{tab:potentialsiv}, respectively.

Note also that the CSB potential strengths,
being very small at larger values of $\sigma$, grow even more
rapidly as the range becomes shorter; indeed at $\sigma\simeq 0.6$~fm
the CSB corrections become of the same size as the
charge-symmetric potential. This hints at the possibility that
the breakdown scale in the presence of the Coulomb interaction
could be even lower than for pure short-range potentials.

\begin{table}
\caption{Potential strength parameters corresponding to $\sigma=0.6,\ 0.8,\ 1.0,$ and $1.2$~fm, see Eq.~(\ref{eq:vnn}).}
\label{tab:potentials}\begin{tabular}{lrrrrrrr}
\hline
$\sigma$ [fm] & $A_1$ [fm${^{-1}}$]& $A_2$ [fm${^{-1}}$]& $A_3$ [fm${^{-3}}$]& $A_4$ [fm${^{-3}}$]& $A_5$ [fm]& $A_6$ [fm]& $A_7$ [fm${^{-1}}$]\\
\hline
 $0.6$ & $-0.677656 $ & $-0.183947 $ & $0.209219 $ & $0.078202 $ & $-0.967619 $ & $-0.144969 $ & $-0.571874 $ \\
 $0.8$ & $-1.421301 $ & $-0.068006 $ & $0.425644 $ & $0.017542 $ & $-0.355897 $ & $-0.044137 $ & $-0.140770 $ \\
 $1.0$ & $-0.776962 $ & $-0.006286 $ & $0.153822 $ & $-0.000171 $ & $-0.173443 $ & $-0.022155 $ & $-0.055884 $ \\
 $1.2$ & $-0.470719 $ & $0.005275 $ & $0.066913 $ & $-0.001572 $ & $-0.097818 $ & $-0.015785 $ & $-0.029482 $ \\
\hline
\end{tabular}
\end{table}

\begin{table}
\caption{CSB potential strength parameters corresponding to $\sigma=0.6,\ 0.8,\ 1.0,$ and $1.2$~fm, see Eq.~(\ref{eq:vnniv}).}
\label{tab:potentialsiv}\begin{tabular}{lrrr}
\hline
$\sigma$ [fm] & $A_1^\mathrm{CSB}$ [fm${^{-1}}$]& $A_3^\mathrm{CSB}$ [fm${^{-3}}$]& $A_5^\mathrm{CSB}$ [fm]\\
\hline
 $0.6$ & $ 1.433500 $ & $-0.792231 $ & $-0.128230$ \\
 $0.8$ & $ 0.018066 $ & $-0.006140 $ & $-0.042043$  \\
 $1.0$ & $-0.000549 $ & $-0.000520 $ & $-0.017543$  \\
 $1.2$ & $ 0.002286 $ & $-0.000815 $ & $-0.009687$  \\
\hline
\end{tabular}
\end{table}

\begin{figure}
\includegraphics[width=\textwidth]{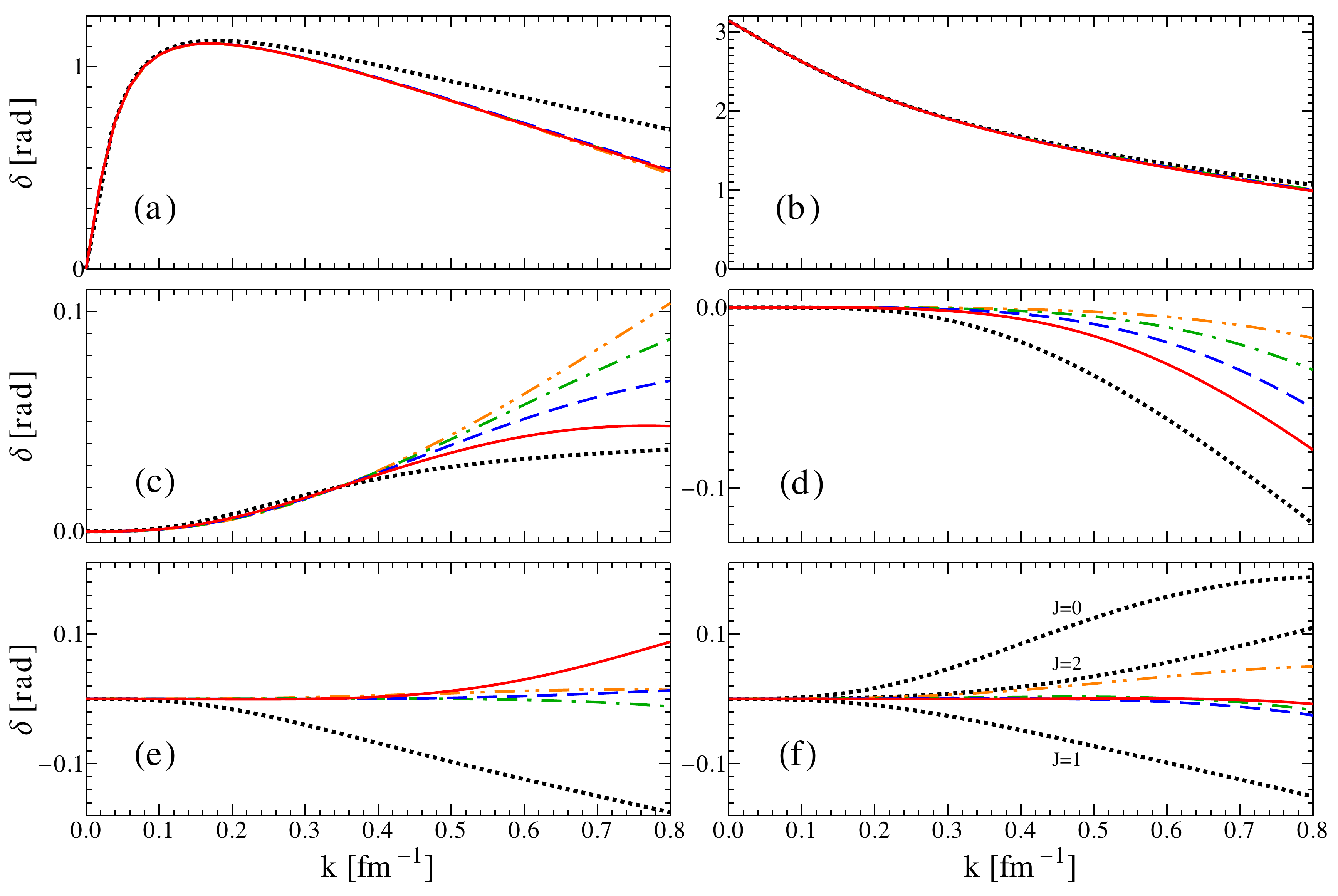}
\caption{(Color online) Low-energy $pn$ phase shifts as described by our $NN$ potentials.
The panels show the phase shifts as follows:
(a) ${^1}S_0$; (b) ${^3}S_1$; (c) $\epsilon_1$; (d) ${^3}D_1$;
(e) ${^1}P_1$; (f) ${^3}P_{0,1,2}$.
Curves corresponding to the different
values of $\sigma$ are: red solid: $\sigma=1.2~\mathrm{fm}$,
blue dashed: $\sigma=1.0~\mathrm{fm}$,
green dash-dotted: $\sigma=0.8~\mathrm{fm}$,
orange double-dot-dashed: $\sigma=0.6~\mathrm{fm}$.
The black dotted line shows the results of the Nijmegen PWA93 
analysis~\cite{Stoks:1993tb,nn-online}. 
Note that our calculation gives identical results for the 
${^3}P_{0,1,2}$ phase shifts, see panel (f); 
the PWA93 curves for these phase shifts are labelled in the figure.}
\label{fig:phaseshifts}
\end{figure}

\begin{figure}
\includegraphics[width=\textwidth]{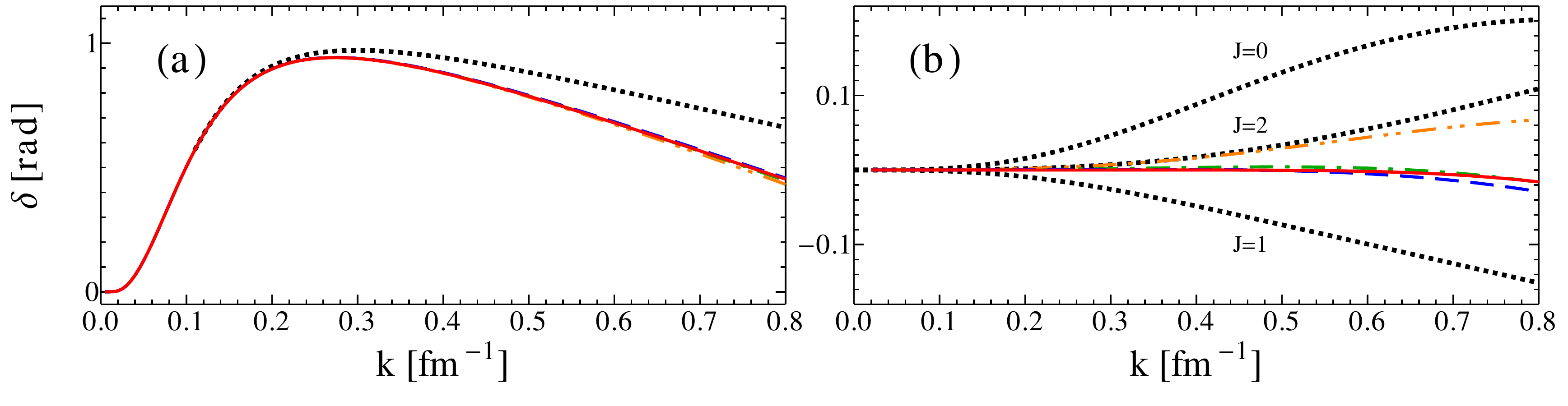}
\caption{(Color online) Low-energy $pp$ phase shifts as described by our $NN$ potentials.
The panels show the phase shifts as follows: (a) ${^1}S_0$; (b) ${^3}P_{0,1,2}$.
The curves are denoted as in Fig.~\ref{fig:phaseshifts}.}
\label{fig:phaseshifts_pp}
\end{figure}

In Figs.~\ref{fig:phaseshifts} and~\ref{fig:phaseshifts_pp}
we show how closely 
the corresponding effective potentials reproduce the 
nucleon-nucleon phase shifts. The $^3S_1$ phase shift 
is reproduced very well by all four potentials.
The $^1S_0$ phase shift is described well by our fits at low momenta,
$k\lesssim 0.35~\mathrm{fm}^{-1}$, both in the $pn$ and
in the $pp$ channel, while at larger momenta
they start to deviate noticeably from the empirical values. 
The $S-D$ mixing angle is also reproduced rather well over
this momentum range. We also show the ${^3}D_1$ phase shift and the
$P$-wave ones (note that to this order the results for the three triplet $P$-wave
phase shifts are identical, as described above). The constraints 
imposed on the $P$-waves mean that their phase shifts are close to zero 
for $k\lesssim 0.35~\mathrm{fm}^{-1}$, and remain much smaller than the
empirical ones across the low-momentum region. The ${^3}D_1$ phase shift,
which is not constrained, also remains smaller than the (already small)
empirical shifts. Note that the plots show phases on a much bigger scale
than the range of applicability of our EFT.

\begin{table}
\caption{Deuteron parameters resulting from the potentials with $\sigma=0.6,\ 0.8,\ 1.0,$ and $1.2$~fm, compared to
experimental values.}
\begin{tabular}{cccc}
\hline
Range $\sigma$ [fm] & Energy [MeV] & Charge radius [fm]\\
\hline
 $0.6$ & $-2.207$ & $2.123$  \\
 $0.8$ & $-2.207$ & $2.120$ \\
 $1.0$ & $-2.204$ & $2.118$ \\
 $1.2$ & $-2.198$ & $2.116$ \\
\hline
 exp.  & $-2.224$ & $2.130\pm0.010$~\cite{Sick} \\
 \hline
\end{tabular}
\label{tab:deuteron}
\end{table}

Table~\ref{tab:deuteron} shows the deuteron energy and charge radius
resulting from the four selected potentials. (See Sec.~\ref{sec:charge_radii} for a discussion of nuclear charge radii.)
As expected from the fact that these potentials describe the low-energy 
triplet scattering parameters very well, the results for the 
deuteron are very close to the experimental values.
Note that the tensor potential, which is responsible for the 
$S-D$ mixing, is crucial for the deuteron binding (despite
the mixing angle $\epsilon_1$ being very small). The bound state does 
not survive if this potential is set to zero.

\section{Three and four nucleons}
\label{sec:NNN}

Having determined the parameters of the two-body potentials, 
we now apply them to the three- and four-body nuclei. 
However, to do this, we first need to determine the parameters 
of the three-body potentials to be used in conjunction with them.

\subsection{Ground-state energies}
\label{sec:energies}

\begin{figure}
\includegraphics[width=\textwidth]{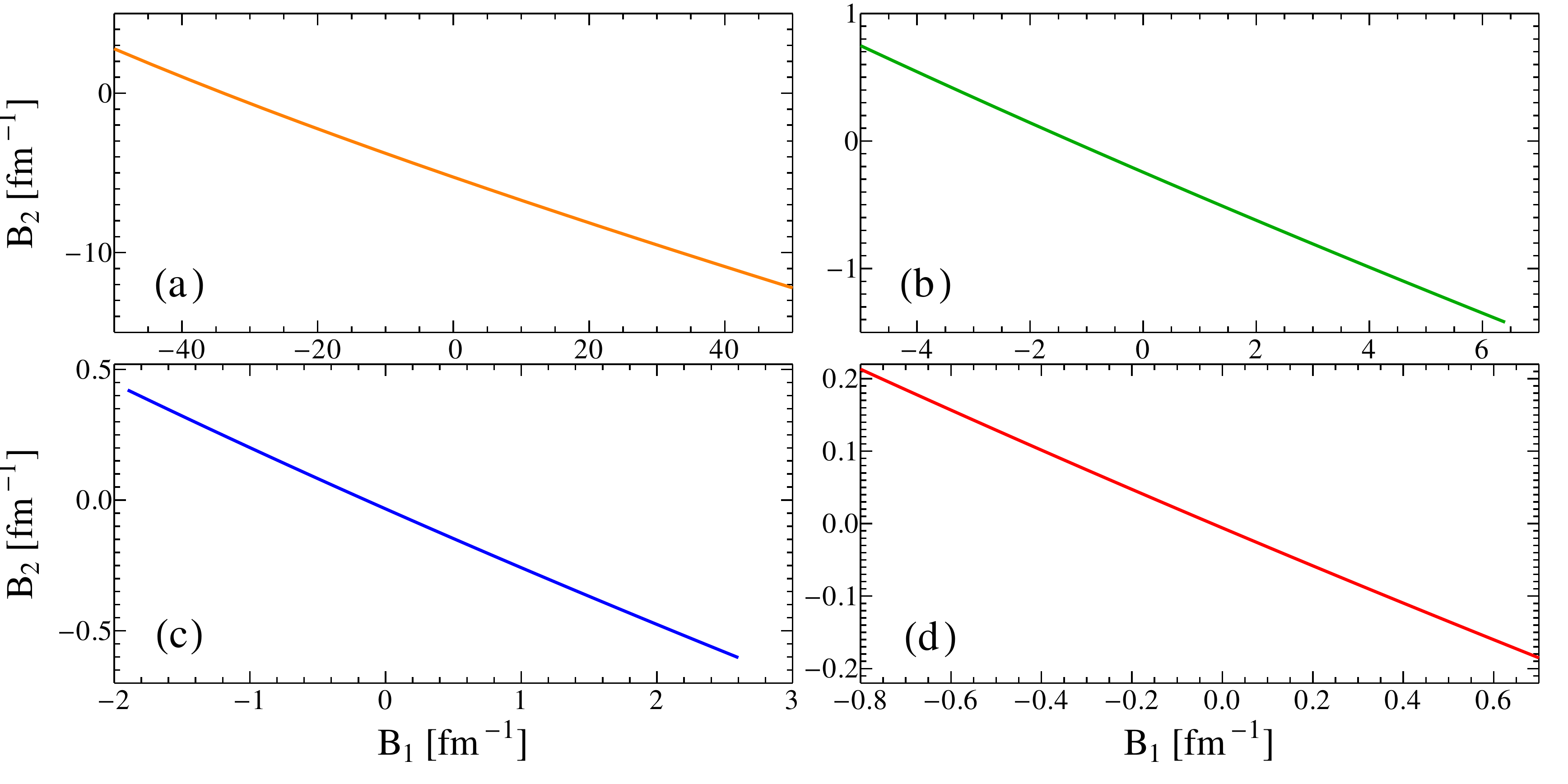}
\caption{(Color online) Graphical representation of the relationship
between the parameters $B_1$ and $B_2$ that 
reproduces the triton binding energy  $E({^3}H)=-8.48$~MeV.
Panels (a)--(d) correspond to $\sigma = 0.6,\ 0.8,\ 1.0$, and $1.2$~fm, in order.
}
\label{fig:3HLONLO}
\end{figure}

Our fitting procedure in the three- and four-nucleon sectors is 
as follows. First, we use the SVM to calculate the triton 
ground-state energy $E({^3}H)$ in an appropriate region of the plane 
of the three-body parameters $B_1$ and $B_2$, Eq.~(\ref{eq:vnnn}).
Then, we identify those pairs of values of $B_1$ and $B_2$ that result 
in an energy that matches its observed value,
$E({^3}H)=-8.48$~MeV. The locus of
these points in the $B_1 B_2$ plane is shown in Fig.~\ref{fig:3HLONLO}.
For each choice of the cut-off $\sigma$, these form almost straight
lines. Another striking feature is the strong dependence 
of the typical sizes of these parameters on $\sigma$,
with those for $\sigma=0.6$~fm being about two orders of magnitude
larger than those for 1.2~fm. This is a signal that for $\sigma=0.6$~fm 
the expansion of our EFT is breaking down. The $\sigma=0.8$~fm potential
is a borderline case: even though it does not show issues similar to
those for the $\sigma=0.6$~fm potential as described below, the
potential strengths show the tendency to become large already at this value of
the cut-off.

\begin{figure}
\includegraphics[width=\textwidth]{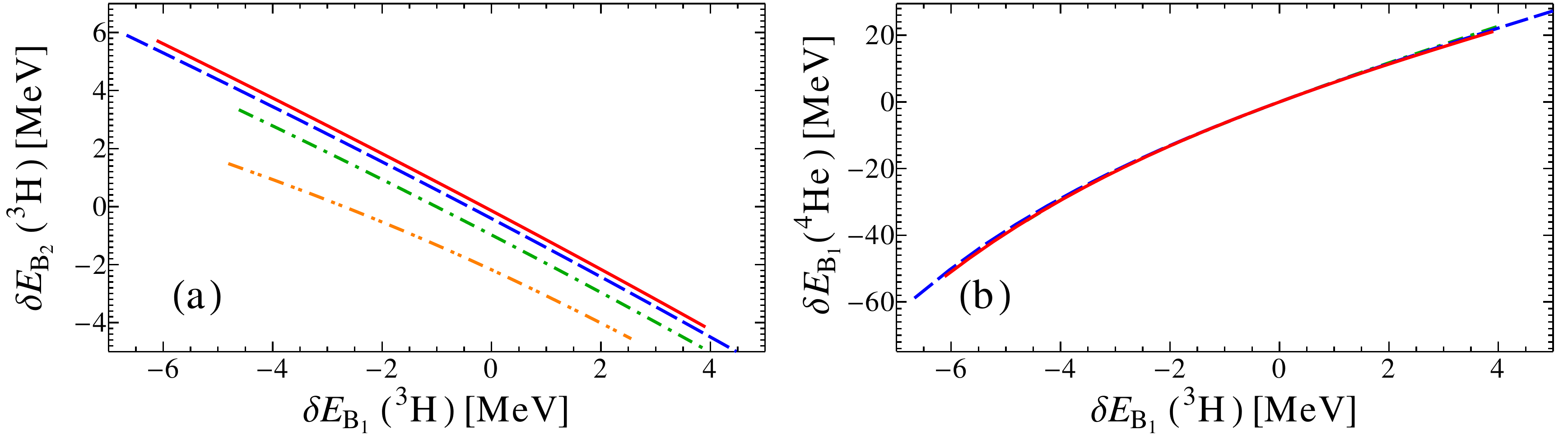}
\caption{(Color online) (a) Correlation between the 
contributions of the two three-nucleon potentials, $B_1$ and $B_2$,
to the triton ground-state energy for the values shown in
Fig.~\ref{fig:3HLONLO}. The behaviour for different values of $\sigma$
is represented by colour and line type as in Fig.~\ref{fig:phaseshifts}.
(b) Correlation between the contributions of $B_1$
to the triton and ${^4}$He ground-state energies. As explained in the main text,
no calculation for $\sigma=0.6$~fm has been performed.
\label{fig:3HLONLOE}
}
\end{figure}

The contributions, $\delta E_{B_1}$ and $\delta E_{B_2}$,  of the two
three-body potentials to the binding energy of $^3$H are shown in panel (a) of Fig.~\ref{fig:3HLONLOE}. It can be seen that, 
as a result of our use of an implicit renormalisation scheme, the
contributions of these two three-body terms are of similar sizes 
for all potentials from the fits shown in Fig.~\ref{fig:3HLONLO}. 
One can also see from Fig.~\ref{fig:3HLONLOE} that
the triton would be under-bound if only two-nucleon forces were included.

Panel (b) of Fig.~\ref{fig:3HLONLOE} shows the contributions of the LO three-nucleon
potential to the $^3$H and $^4$He binding energies. In the absence of 
the NLO potential,
these lines would form the ``Tjon lines'' for our EFT \cite{Tjon:1974}.
Note that they do not yet include the CSB three-body contribution 
in ${^4}$He. 
This figure does not show the $\sigma=0.6$~fm curve;
as can be inferred from Tables~\ref{tab:potentials} and~\ref{tab:potentialsiv}, the
CSB corrections are too big
for the $\sigma=0.6$~fm potential, which expected to lead to troubles already
in ${^3}$He. Furthermore, the large cancellations between the LO and NLO three-body
contributions to ${^3}$H cannot be maintained in ${^4}$He even without the
CSB corrections of Eq.~\eqref{eq:vnniv}. All this clearly
signals that this cut-off is beyond the breakdown scale of the EFT;
we will therefore present no further results for the $\sigma=0.6$~fm potential.

\begin{figure}
\includegraphics[width=0.55\textwidth]{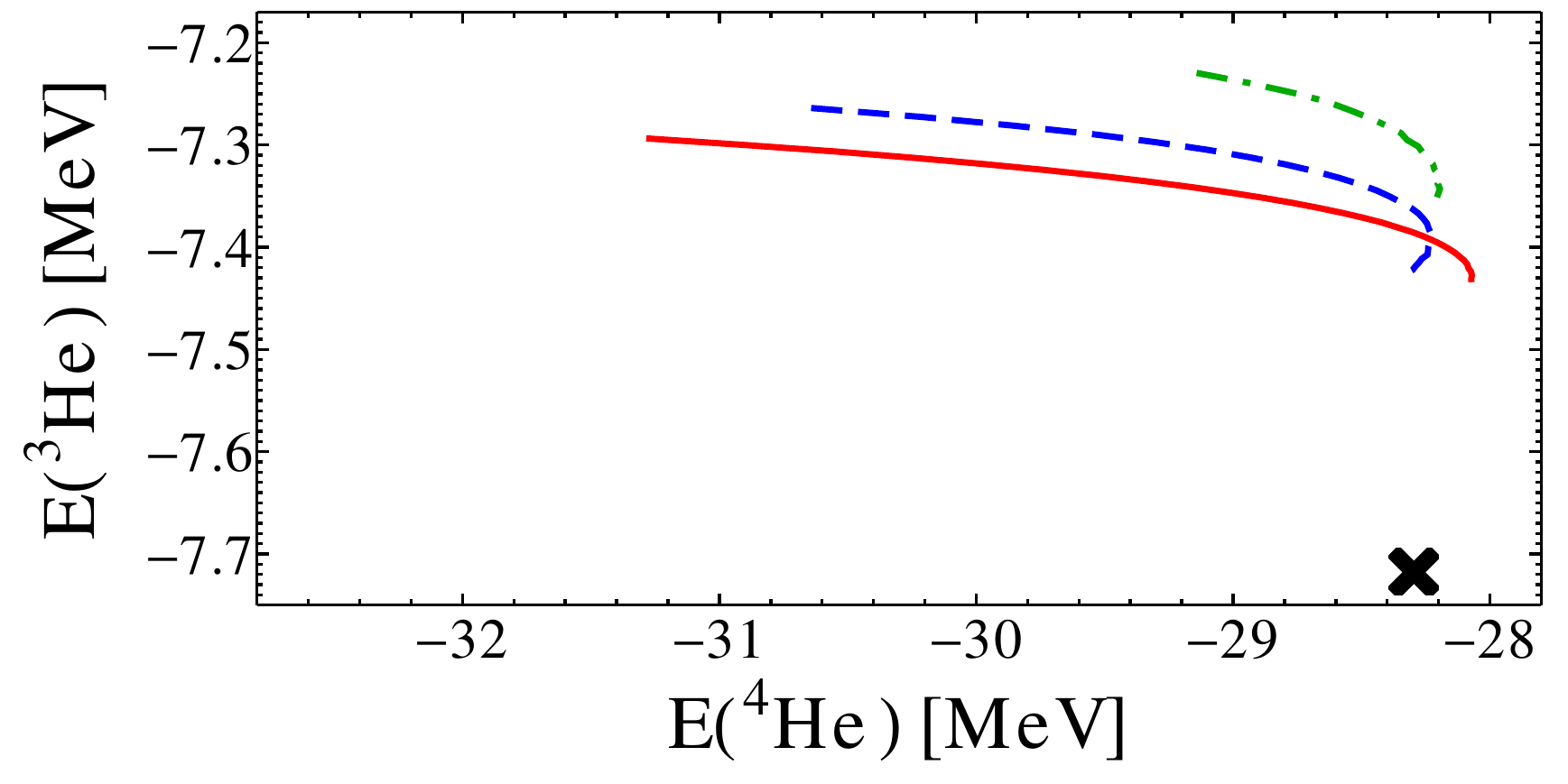}
\caption{(Color online) Correlation between the results for the
ground-state energies of ${^3}$He and ${^4}$He,
calculated without the CSB three-nucleon force. The lines for different values 
of $\sigma$ are denoted as in Fig.~\ref{fig:phaseshifts}. The cross marks the experimental values.
}
\label{fig:He3He4energies}
\end{figure}
Figure \ref{fig:He3He4energies} shows the relationship between the energies
of the two helium isotopes obtained from our fitted potential, 
at this stage including the Coulomb and CSB nucleon-nucleon potentials
but not yet the CSB three-body interaction. One can see that ${^3}$He
binding energy is underestimated, whereas that of ${^4}$He tends to be overestimated
along most of the length of the curves (even though the latter reach the
experimental value at one end).
The under-binding of $^3$He is rather small ($\sim 5\%$) but it does not
depend strongly on the choice
of three-body fit. In contrast the $^4$He energy varies by up to $10\%$ as
the parameters are varied over the ranges shown in Fig.~\ref{fig:3HLONLO}.
The spread between the curves for the different values of $\sigma$
can be taken as an estimate of the uncertainties in our results due to
omission of higher-order contributions. This is discussed further in
Appendix B, which also 
contains a discussion of the numerical convergence of
the SVM.

Finally we add the CSB three-nucleon interaction, fitting it to 
the observed
${^3}$He ground-state energy, $E({^3}\mathrm{He})=-7.718$~MeV.
It is worth pointing out that this potential has a small effect
on observables, in particular, the binding energy,
and so, in principle, its contributions could be calculated
perturbatively, by appropriately scaling the value of $B_1$
in the charge-symmetric ${^3}$He calculation:
\begin{equation}
\delta E_\mathrm{CSB}({^3}\mathrm{He})=\eta_\mathrm{CSB}\,
\delta E_{B_1}({^3}\mathrm{He})\,,
\end{equation}
where $\eta_\mathrm{CSB}$ is fitted to give the observed energy of
${^3}$He. With this value for $\eta_\mathrm{CSB}$, the contribution of 
the CSB interaction to ${^4}$He can also be estimated in the same spirit,
from
\begin{equation}
\delta E_\mathrm{CSB}({^4}\mathrm{He})=\frac{1}{2}\eta_\mathrm{CSB}\,\delta E_{B_1}({^4}\mathrm{He})\,,
\end{equation}
where the factor $1/2$ takes into account the symmetry of the ${^4}$He 
wave function with respect to interchange between protons and neutrons.
This symmetry is only approximate since the Coulomb potential already
leads to CSB at the two-nucleon level. For a precise determination of 
three-body CSB effects, we have used the SVM to solve Schr\"odinger
equation with both the Coulomb potential and the CSB three-body force,
using the results for $^3$He to determine the three-body strength $B_\mathrm{CSB}$.
The results are within a few percent of the perturbative estimate
(from $2\%$ to $4\%$ for $\sigma = 1.2$~fm 
or $6\%$ to $9\%$ for $\sigma=0.8$~fm).

\begin{figure}
\includegraphics[width=\textwidth]{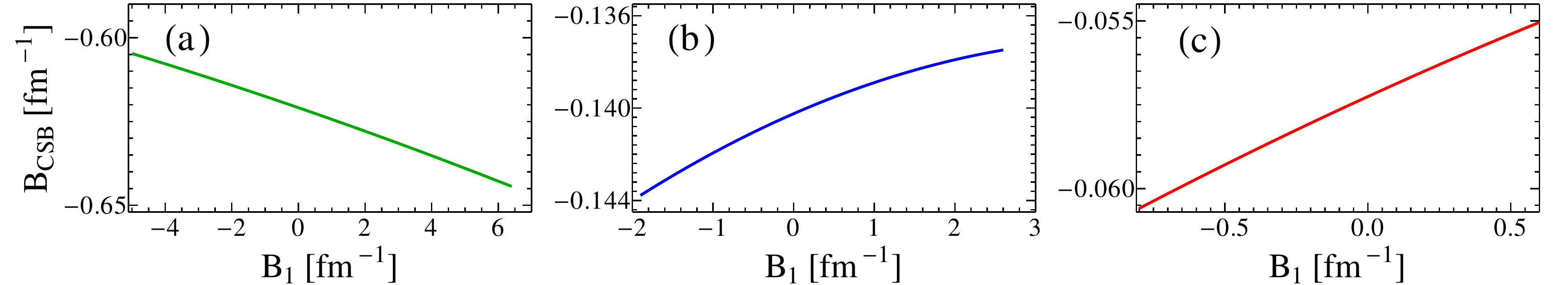}
\caption{(Color online) Correlation of the strength $B_\mathrm{CSB}$ of
the CSB three-nucleon potential with the value of $B_1$, for potentials
fitted to the binding energies of $^3$H and $^3$He. Panels (a)--(c) correspond
to $\sigma=0.8$, $1.0$, and $1.2$~fm, in order.
}
\label{fig:CSB3Npot}
\end{figure}
Fitting the ground-state energies of $^3$H and $^3$He leads to a 
one-parameter family of three-body potentials. The relationship between
the strength of the CSB three-nucleon potential and that of the LO
charge-symmetric potential, $B_1$, is shown in Fig.~\ref{fig:CSB3Npot}. 
For $\sigma=0.8$~fm and $\sigma=1.2$~fm, the correlation line is again
indistinguishable from linear. For $\sigma=1.0$~fm, however, one can
notice a visible deviation from straight line. This can be explained by
noting that the $\sigma=0.8$~fm and the $\sigma=1.2$~fm lines have slopes
of different signs; the transition from the negative to the positive slope
must thus take place at some value of $\sigma$ between those two values.
The $\sigma=1.0$~fm value is apparently close to that transition, therefore
higher-order (quadratic) effects become visible in the corresponding curve.

This fit to both three-nucleon energies leaves only the binding energy of 
$^4$He as a prediction of our potentials. Including the CSB force removes
the under-binding of ${^3}$He. At the same time it makes ${^4}$He
over-bound for all our potentials: a simple combinatorial argument gives
that the shift of about $-0.5$~MeV needed to correct for the under-binding
of ${^3}$He translates into a shift of about $-1$~MeV in the energy of
${^4}$He; the latter shift will as a rule be even bigger since ${^4}$He 
is a denser system. However a single binding energy can provide
only limited information on how well our EFT is describing 
the ground states of these light nuclei and so we now turn to
calculations of their charge radii.

\subsection{Charge radii}
\label{sec:charge_radii}

The SVM provides wave functions for the ground states of the three- 
and four-nucleon systems, as well as their binding energies. 
To test how well our EFT describes the structure of these states, 
we have used these wave functions to calculate charge radii. 
These are accessible experimentally through
electron scattering~\cite{Sick} or isotopic shift measurements.
They can be calculated by expanding the charge form factor
\begin{equation}
F_C(q^2) = 1-\frac{q^2 r^2}{6}+\dots\,
\end{equation}
where the $F_C(q^2)$ are defined in terms of the spin-averaged matrix
elements of the EM density by \cite{NED}
\begin{equation}
F_C(q^2)=\frac{1}{Z}\bra{+\svec q/2}J^0_\mathrm{em}(\svec q)
\ket{-\svec q/2}\,.
\end{equation}

The lowest-order coupling of the photon to the nucleon charge contributes 
to the form factor at $\mathcal{O}(Q^0)$.
To the order we work here, we also need operators that contribute
to the form factor at two orders higher, $\mathcal{O}(Q^2)$. 
The most important of these contain low-energy constants related to 
the charge radii of the nucleons. Two-body contributions to the charge 
radii appear first at $\mathcal{O}(Q^3)$
\cite{Valderrama:2014vra}, which is beyond the order considered here.
As a result, we can write
\begin{equation}
r^2 = \frac{1}{Z}\bra{\Psi_0}\sum\limits_{j=1}^A \frac{1}{2}\left(1+\tau_3\right)_j\svec{r}_j^2\ket{\Psi_0}+r_p^2 + \frac{3}{4M^2} + \frac{N}{Z}r_n^2\,,
\label{eq:chr}
\end{equation}
where $\ket{\Psi_0}$ denotes the internal wave function of the ground state of the nucleus with $Z$ protons and $N=A-Z$ neutrons. 
Here $r_p = 0.8751$~fm is the proton charge radius~\cite{CODATA}, 
$r_n^2 = -0.1161~\mathrm{fm}^2$ is the neutron charge radius 
squared~\cite{Kopecky:1995zz,Kopecky:1997rw}, 
and $3/(4M^2)$ is the Foldy correction, which is omitted from
the atomic-physics definition of the proton radius \cite{Friar:1997js}.

In principle there are further relativistic recoil terms in the 
photon-nucleon vertex and in the nucleon propagators that are also 
of order $Q^2$. However, for the case of the deuteron it is known that 
these leading relativistic corrections to the charge radius are 
numerically very small~\cite{Chen:1999tn}.
On dimensional grounds, the corrections to Eq.~\eqref{eq:chr} must scale 
as $C/M^2$, where $C$ is a constant. If one takes $C$ to be of the order 
of one (and it is much smaller than that for the deuteron), these 
corrections result in shifts of less than $\pm0.01$~fm in the 
values of the charge radii. We have also calculated some of these corrections
for terms that stem from $1/M^2$ recoil corrections to the 
photon-nucleon vertex. We find that these make very small contributions 
to the nuclear charge radii, about $0.003$~fm for  ${^4}$He and even less 
for the three-nucleon systems.

\begin{figure}
\includegraphics[width=\textwidth]{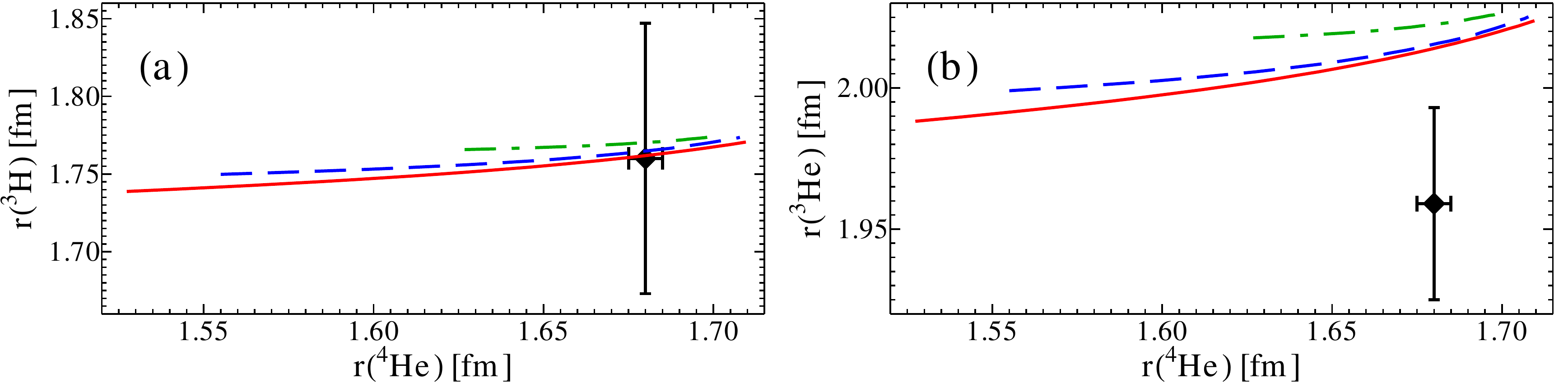}
\caption{
%\comment{CODATA $r_p$  } 
(Color online) (a): Correlation between ${^4}$He and 
${^3}$H charge radii; (b): Correlation between ${^4}$He and 
${^3}$He charge radii. The lines for different values 
of $\sigma$ are denoted as in Fig.~\ref{fig:phaseshifts}.
The experimental values are taken from Ref.~\cite{Sick}.
}
\label{fig:chr34}
\end{figure}

With the wave functions from the SVM, it is straightforward to evaluate 
Eq.~\eqref{eq:chr} for the charge radii. 
The results are shown in Table~\ref{tab:deuteron} for the deuteron 
and in Fig.~\ref{fig:chr34} for ${^3}$H, ${^3}$He, and ${^4}$He.
In the case of the deuteron, they agree with the experimental value to better 
than $1\%$ for all four choices of regulator.

The results in Fig.~\ref{fig:chr34} show that all families
of three-body potentials are in remarkably good 
agreement with the experimental value for the $^3$H charge radius.
The values for the charge radius of ${^3}$He are somewhat larger than 
the experimental one, but the discrepancy is never more than two 
standard deviations. The significance of the differences is further 
reduced if we take account of the theoretical uncertainties in our
results, as indicated by the spread between the curves. 
It may also be worth noting that taking the value for the proton
charge radius obtained from muonic hydrogen~\cite{Antognini:1900ns},
$r_p=0.8409$~fm, would shift the ${^3}$He results down to within $1.5$
standard deviations of the experimental value.
Another interesting feature is that the three-nucleon
CSB force that compensates for the under-binding of ${^3}$He is also
quite important for the description of the ${^3}$He charge radius; without
this force we would get larger radii (as expected), the typical values being
about $2.05$~fm.

\begin{figure}%[ht]
\includegraphics[width=0.5\textwidth]{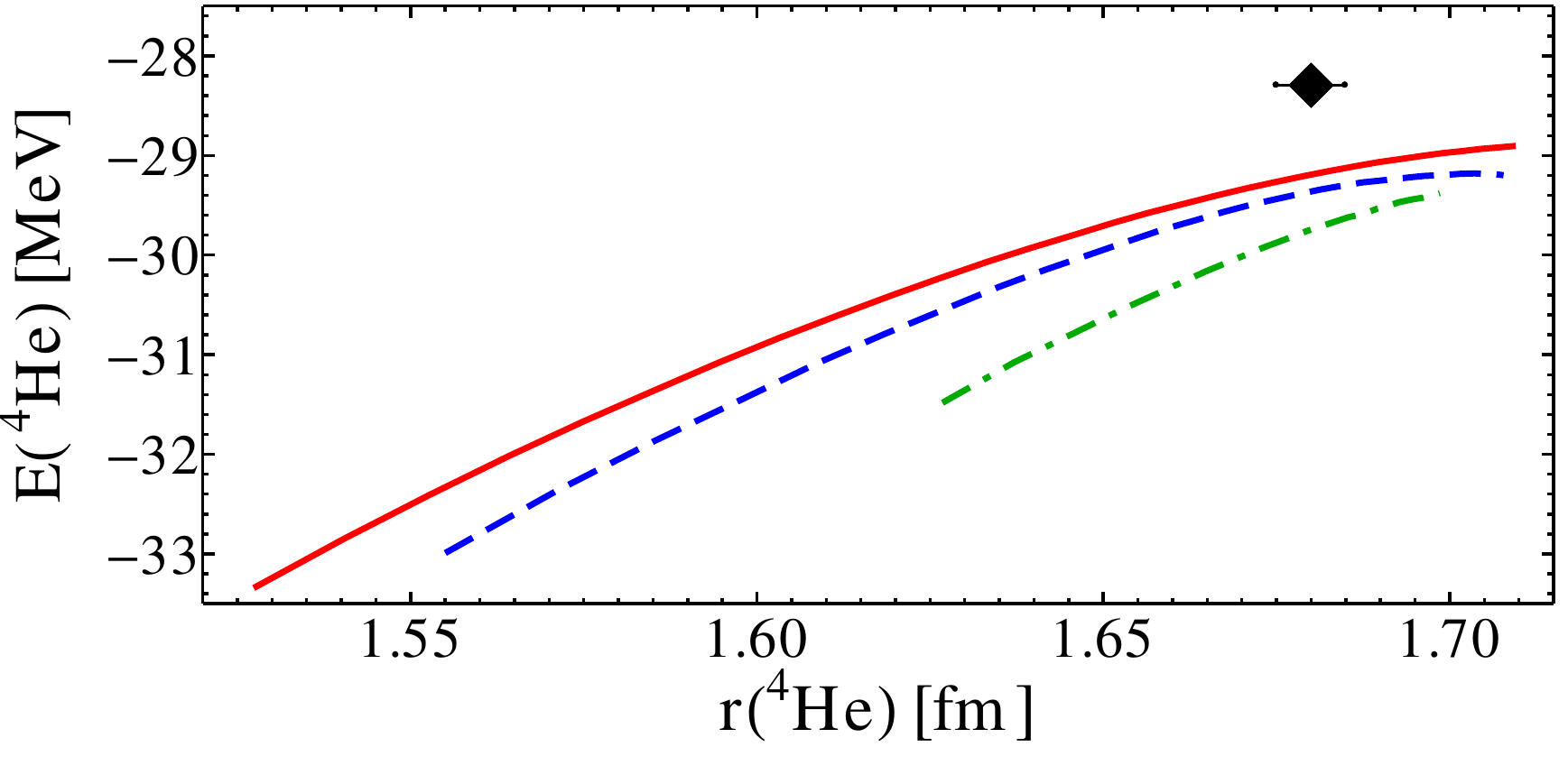}
\caption{
%\comment{CODATA $r_p$  } 
(Color online) Correlation between ${^4}$He charge
radius and ground state energy. The notation 
is as in Fig.~\ref{fig:3HLONLO}. 
}
\label{fig:chre4}
\end{figure}
Finally, in Fig.~\ref{fig:chre4}, we present our
results for the energy and the charge radius of $^4$He 
for our families of three-body potentials. For each
family, the charge radius shows the expected correlation with 
energy, decreasing with increasing binding energy, as
expected. As already mentioned, $^4$He is still over-bound, 
by about $1..1.5$~MeV for potentials that reproduce the observed
charge radius. This difference is small 
(about $3\%$ to $5~\%$ of the binding energy) 
and within the expected truncation error of the EFT expansion.

As discussed further in Appendix B, we estimate the
uncertainty from the spread of the results for different
choices of cut-off. For the potentials that describe all
four observables, this spread is typically about $3\%$,
which is consistent with the expected error due to
our truncation of the EFT at NNLO, which is $\sim (1/3)^3$
for an expansion parameter which we estimate to be $Q\simeq 1/3$.

\section{Discussion\label{Sec:discussion}}
In this work we have studied nuclei with $A=3$ and $4$ in the framework 
of the pionless EFT, working at NNLO. We use a Hamiltonian formulation 
rather than the more usual Lagrangian version, expressing the 
interactions in the form of local potentials with a Gaussian regulator. 
As well as short-range two- and three-body forces, we also include 
the Coulomb potential. To the order we work, this also requires that 
we add CSB two- and three-body contact interactions.

The resulting Schr\"odinger equation is solved variationally using the 
SVM with a basis of Gaussian trial functions. This provides ground-state
energies and wave functions for the nuclei with $A=3$ and $4$. Since our 
three-body potential contains three terms, fitting the energies of 
$^3$H and $^3$He leaves us with a one-parameter family of potential for 
each choice of cut-off.

In this approach, the whole potential is treated to all orders. This 
means that we cannot renormalise it perturbatively; instead we use an 
implicit renormalisation. If the regulator scale is taken above the 
breakdown scale of the EFT, this can lead to very large values for
the parameters with extreme fine-tuning in order to reproduce observables.
We find that the determination of parameters becomes unstable for 
potential with ranges $\sigma < 0.8\text{ fm}$. Moreover, 
the higher-order effects of these fine-tuned potentials are large for 
nuclei with $A>2$, and thus the results for binding energies are actually 
quite poor. We therefore reject these unnatural solutions.

For each of these potentials we have calculated the binding energy of 
$^4$He and the charge radii of the nuclei with $A=3$ and $4$. 
The dependence of the results on the cut-off provides a measure of the
uncertainties in our results. These are about $3\%$, which matches what we expect for our truncation of the EFT. We see some indications that the
agreement with empirical data for $A=4$ is slightly poorer than for $A=3$, which is not a
surprise given that $^4$He is a denser system, with higher typical momenta.

Within these uncertainties and those of the data, we are able to find
potentials that give good agreement with the measured values of
the $^4$He energy and the charge radii. 

In this paper we have concentrated on the specific form of the interaction
that applies to the (almost) Wigner-$\mathrm{SU}(4)$-symmetric $A\leq4$ nuclei.
Extending our method to mixed-symmetry nuclei with $A>4$ will require
determination of many more parameters. A natural way to do this
is through fitting scattering observables. This will require an
extension of SVM using, for example, the Kohn variational principle.

\section*{Acknowledgements}

This work was supported by the UK STFC under Grants ST/F012047/1 
and ST/J000159/1, by the Deutsche Forschungsgemeinschaft (DFG) through
the Collaborative Research Center ``The Low-Energy Frontier of the
Standard Model'' (SFB 1044), by the Russian Foundation for Basic Research
Grant No.\ NSh-3830.2014.2, and by the Moscow Engineering Physics Institute
Academic Excellence Project (Contract No.\ 02.a03.21.0005). 
The authors would like to acknowledge the assistance given by IT Services and the use of the Computational Shared Facility at The University of Manchester.
We are grateful to
H.~Grie{\ss}hammer, J.~Kirscher, D.~Phillips, E.~Epelbaum,
and V.~Baru for helpful comments and discussions.

\appendix
\section{Kohn variational principle\label{sec:Kohn}}
In order to construct a low-energy nucleon-nucleon potential, we employ the complex variant of the Kohn variational method
(see Refs.~\cite{Miller1987,McCurdy1987,Schneider:1988zz}).
The idea of this method is based on the fact that the Lippmann-Schwinger equation
for the scattering operator $T$,
\begin{equation}
T(E) = V + V G^{(+)}(E) V\,,
\label{eq:Toperator}
\end{equation}
can be recast into the problem of finding the stationary point of a  functional involving
either $T$ or the exact Green's function with the outgoing-wave (radiation) boundary condition,
\begin{equation}
G^{(+)}(E)=\left[E-H+i0\right]^{-1}\,.
\end{equation}
For a particular choice of the trial wave functions
(see the above cited references for
further details of the derivation which we omit),
the resulting variational expression for the Green's function projected onto partial waves
reads:
\begin{equation}
G^{(+)}(r',r)=\sum\limits_{i,j}\phi_i(r') \left[\mat{E}-\mat{H}\right]^{-1}_{ij} \phi_j(r)\,,
\label{eq:kohngreenfunction}
\end{equation}
where the matrix $\left[\mat{E}-\mat{H}\right]^{-1}$ is the inverse of the square matrix with entries $\left\langle\phi_i^*\right|\left[E-H\right]\left|\phi_j\right\rangle$.
The radiation condition is carried by one of the wave functions, and the remainder remains finite and is square-integrable,
\begin{equation}
\begin{split}
 &\phi_0(r)\propto \frac{e^{i k r}}{r},\quad r\to \infty\,,\\
 &\phi_j(r)\to 0, \quad r\to \infty\,,\ j>0\,,\\
 &\phi_j(r)\propto r^l,\quad r\to 0\,,\ j>0\,,
\end{split}
\label{eq:kohnasymptotic}
\end{equation}
with $k=\sqrt{M E}$ being the relative momentum and $l$ the angular momentum in the corresponding partial wave. 
Note that the vectors $\phi(r')$ in Eq.~\eqref{eq:kohngreenfunction} are not complex conjugated as one would naively expect; 
this follows from the radiation condition on the Green's function~\cite{Miller1987}.

As long as the basis functions satisfy the conditions given by Eq.~\ref{eq:kohnasymptotic},
one can select them based on convenience of use in the particular calculation.
The basis used by us is given by
\begin{equation}
\begin{split}
&r \phi_0(r)=i k^{l+1}r \left(1-e^{-r/\alpha_0}\right)^{2l+1}h^{(+)}_l(kr)\,,\\
&r \phi_j(r)=\sqrt{\frac{2(j-1)!}{\alpha(j+2l-1)!}}\left(2r/\alpha\right)^{l+1} e^{-r/\alpha} L^{(2l+2)}_{j-1}(2r/\alpha)\,,\ j>0\,,
\label{eq:kohnbasis}
\end{split}
\end{equation}
where $h^{(+)}_l(x)$ is the spherical Hankel function of the first kind, $L^{(2l+2)}_j(x)$ are the associated Laguerre polynomials,
and $\alpha_0$ and $\alpha$ are real parameters that can be fine-tuned so as to improve the convergence of the calculation.
The results shown in this work were obtained with $\alpha_0=1$~fm, $\alpha=0.25$~fm.
The free solutions to the wave equation are  the spherical Bessel functions,
\begin{equation}
\psi_l(k,r)=j_l(k r)\,. 
\label{eq:jlkr}
\end{equation}
With this normalisation, the matrix elements of the scattering operator (\ref{eq:Toperator}) are just minus the scattering amplitude
between the corresponding partial waves:
\begin{equation}
f_{ll'}(k)=-\left\langle\psi_l\right|V\left|\psi_{l'}\right\rangle-
\left\langle\psi_l\right|VG^{(+)}(E)V\left|\psi_{l'}\right\rangle\,.
\label{eq:fll}
\end{equation}
Equations (\ref{eq:kohnbasis}--\ref{eq:fll}) can be used to extract the nucleon-nucleon scattering parameters such as scattering
lengths and effective ranges resulting from the specific potential $V$, and thus allow for
fitting the parameters of the potential. 

Proton-proton scattering can also be treated by this method, using a
distorted-wave formulation~\cite{Miller1987}. Namely, the radiation
condition of Eq.~\eqref{eq:kohnasymptotic} needs to be replaced by 
the Coulomb radiation condition:
\begin{equation}
\begin{split}
 &\phi_0(r)\propto \frac{e^{i\left( k r - \eta \ln 2 k r+\delta_l\right)}}{r},\quad r\to \infty\,,\\
\end{split}
\label{eq:kohnasymptoticcoulomb}
\end{equation}
where
\begin{equation}
\eta=\frac{M\alpha_\mathrm{em}}{2k},\quad \delta_l=\Arg\Gamma(1+l+i\eta)\,.
\end{equation}
The Coulomb-modified basis functions that we use are
\begin{equation}
\begin{split}
&r \phi_0(r)=(-i k)^{l}e^{\pi \eta/2 + i\delta_l} \left(1-e^{-r/\alpha_0}\right)^{2l+1}W_{-i\eta,l+1/2}(-2ikr)\,,\\
\label{eq:kohnbasiscoulomb}
\end{split}
\end{equation}
with $\phi_j(r)$ given by Eq.~\eqref{eq:kohnbasis} for $j>0$, with the 
same values of $\alpha_0$ and $\alpha$. The
free spherical Bessel functions, in turn, are replaced by the
regular solutions to the Coulomb equation:
\begin{equation}
\psi^C_l(k,r)=\frac{C_l}{kr}(-i/2)^{l + 1} M_{i\eta, l + 1/2}(2 i k r)\,, 
\label{eq:jlkrcoulomb}
\end{equation}
where
\begin{equation}
C_l=\frac{2^le^{-\pi\eta/2}\left|\Gamma(1+l+i\eta)\right|}{(2l+1)!}\,.
\end{equation}
Here, $W_{a,b}(z)$ and $M_{a,b}(z)$ are the Whittaker functions
(for details see, e.g., Refs.~\cite[Chapters 13, 33]{dlmf}
and~\cite[Appendix B]{Messiah}).
With these definitions, the strong scattering amplitude
in the presence of the Coulomb interaction is given by
\begin{equation}
f_{ll'}(k)=-\left\langle\psi^C_l\right|V\left|\psi^C_{l'}\right\rangle-
\left\langle\psi^C_l\right|VG^{(+)}(E)V\left|\psi^C_{l'}\right\rangle\,,
\label{eq:fllcoulomb}
\end{equation}
where $V$ is now only the strong part of the potential, whereas $G^{(+)}(E)$
is the full Green's function constructed as given by Eq.~\eqref{eq:kohngreenfunction}
with the full (strong $+$ Coulomb) potential.

\section{Convergence}
\label{sec:convergence}
\subsection{Numerical convergence of the SVM}
\label{sec:convergence:svm}

As a variational method, the SVM gives upper bounds for the
energy eigenvalues, with each consecutive expansion of the basis,
i.e., addition of a new trial function, lowering the energies
\cite{Suzuki:1998bn}.  After having diagonalised the Hamiltonian for $m$ basis states,
 the $(m+1)^{\mathrm{th}}$ trial state
is selected without solving the complete generalised 
eigenvalue problem~\eqref{eq:gevp}.
Only once this state is selected,
the generalised eigenvalue problem
is solved in the $m+1$-dimensional basis.
When the basis size is very large, the energy gain achieved
thanks to expanding it further from $m$ to $m+1$ functions
can become smaller than the numerical error of the solution of the
eigenvalue problem, usually indicating no further gain can be made.

We find that convergence with respect to the size of the basis 
is typically faster for smoother (longer-range)
potentials, as well as for smaller systems. This can also be seen in the
saturation of the basis mentioned above; this also occurs
faster in smoother potentials and in smaller systems.
Table~\ref{tab:basessizes}
lists the bases sizes $m_\mathrm{max}$ used in this work, together with an 
estimate of the change in energy achievable by extending the basis size to
infinity. These estimates are obtained by fitting the tail of the dependence
of the lowest eigenvalue of~\eqref{eq:gevp} by a linear function in the reciprocal basis size $1/m$,
$E_0(m) = E_0^\mathrm{lim}+\Delta E_0/m$. We used the values of $E_0(m)$ with $m>250$
for $A=3$ systems, and $m>400$  for $A=4$.
The uncertainty estimate is then given by 
$\delta E_0=\left|E_0^\mathrm{lim}-E_0(m_\mathrm{max})\right|$. 
The quoted values are the maximal uncertainties for each potential.

\begin{table}
\caption{Basis sizes for $A=3$ and $4$ calculations used in this work
and estimated uncertainties of the ground state energies.
\label{tab:basessizes}}
\begin{tabular}{l@{\hskip 1em}rr@{\hskip 3em}rr}
\hline
\multicolumn{1}{c}{~}& \multicolumn{2}{c@{\hskip 3em}}{${^3}$H, ${^3}$He}& \multicolumn{2}{c}{${^4}$He}\\
\hline
$\sigma$ [fm]& $m_\mathrm{max}$ & $\delta E_0$[MeV] &  $m_\mathrm{max}$ & $\delta E_0$[MeV] \\
\hline
$0.8$& $500$ & $0.03$ & $1200$ & $0.11$\\
$1.0$& $500$ & $0.03$ & $750$ & $0.18$\\
$1.2$& $450$ & $0.02$ & $750$ & $0.07$\\
\hline
\end{tabular}
\end{table}

\subsection{Convergence of the EFT expansion}
\label{sec:convergence:eft}

The method employed in this paper makes it harder to judge the convergence
of the EFT series than it would be in a treatment based on a strict
perturbation theory approach. The use of finite cut-off to regularise the
Schr{\"o}dinger equation, together with implicit renormalisation, means that
there is no clean separation between orders, powers of $Q$, at the level of the potential.
Nevertheless, one would expect that the changes in observables
decrease as one includes terms that contain higher and higher
pieces in the expansion of the potential. 
In addition, the dependence of the results on the regulator
parameter $\sigma$, such as that in Figs.~\ref{fig:chr34}--\ref{fig:chre4},
can provide an estimate of the size of higher-order corrections.

In Table~\ref{tab:convergence} we show a breakdown of the energies
of the $A=2$, $3$ and $4$ systems into
contributions of the various terms in the two- and three-nucleon
potentials, Eqs.~\eqref{eq:vnn}-\eqref{eq:vnniv}, \eqref{eq:vnnn}, and
\eqref{eq:vnnncsb}. The first row again illustrates the problems that arise 
when using a regulator with too short a range. Although the $\sigma=0.6$~fm
potential reproduces the nucleon-nucleon phase shifts and the deuteron
binding energy with the same quality as the other three potentials, it 
gets its largest contribution to the binding of the deuteron 
from the tensor interaction. That interaction, although nominally of NNLO, 
is by far the most important at this scale, which clearly shows that
one cannot keep higher-order effects under control when the cut-off 
is too short-ranged.

\begin{table}
\caption{
Contributions to the ground-state energies of the deuteron, ${^3}$H,
${^3}$He, and ${^4}$He, from different components of the interaction
are shown in Table~\ref{tab:convergence}.
The column labelled $E$ contains the ground-energies; the one labelled
$T$, the sums of the corresponding kinetic and Coulomb energies.
The labels $A_{1\dots 7}$,
$B_{1,2}$, and $B_\mathrm{CSB}$ indicate the contributions of those
interactions to the energies.
All quantities are in units of MeV (except $\sigma$). The three-body and
four-body energies are given for the values of three-body parameters 
that reproduce the observed ${^4}$He charge radius ($1.680$~fm). 
\label{tab:convergence}}
\begin{tabular}{c|cccccccccccc}
\multicolumn{13}{c}{Deuteron}\\
\hline
$\sigma$[fm] & $E$   & $T$   & $A_1$  & $A_2$ & $A_3$ & $A_4$ & $A_5$ & $A_6$ & $A_7$  & $B_1$ & $B_2$ & $B_\mathrm{CSB}$\\
\hline
$0.6    $&$-2.21$&$15.41$&$-5.13 $&$4.18 $&$1.72 $&$-1.93$&$9.65 $&$-4.33$&$-21.76$&$  -  $&$  -  $&$     -         $\\
$0.8    $&$-2.21$&$12.90$&$-27.46$&$3.94 $&$11.55$&$-1.42$&$13.28$&$-4.94$&$-10.05$&$  -  $&$  -  $&$     -         $\\
$1.0    $&$-2.20$&$12.18$&$-24.51$&$0.60 $&$8.88 $&$0.03 $&$9.70 $&$-3.71$&$-5.36 $&$  -  $&$  -  $&$     -         $\\
$1.2    $&$-2.20$&$11.80$&$-20.37$&$-0.69$&$6.69 $&$0.47 $&$6.57 $&$-3.17$&$-3.49 $&$  -  $&$  -  $&$     -         $\\
\hline
\multicolumn{13}{c}{${^3}$H}\\
\hline
$0.8    $&$-8.48$&$30.18$&$-85.43$&$6.13 $&$38.14$&$-2.09$&$32.48$&$-8.87$&$-18.04$&$0.39$&$-1.36$&$     0         $\\
$1.0    $&$-8.48$&$28.74$&$-79.59$&$0.89 $&$30.61$&$0.04 $&$26.83$&$-6.26$&$-9.32 $&$1.57$&$-1.99$&$     0         $\\
$1.2    $&$-8.48$&$28.21$&$-67.99$&$-1.00$&$23.40$&$0.62 $&$19.71$&$-5.31$&$-5.96 $&$1.31$&$-1.46$&$     0         $\\
\hline
\multicolumn{13}{c}{${^3}$He}\\
\hline
$0.8    $&$-7.72$&$24.47$&$-63.40$&$ 7.10$&$27.18$&$-2.51$&$28.11$&$-9.40$&$-17.92$&$0.37$&$-1.28$&$-0.44          $\\
$1.0    $&$-7.72$&$22.86$&$-57.39$&$ 1.05$&$21.20$&$0.05 $&$21.40$&$-6.89$&$-9.25 $&$1.50$&$-1.90$&$-0.36          $\\
$1.2    $&$-7.72$&$22.19$&$-48.23$&$-1.20$&$15.94$&$0.79 $&$15.16$&$-5.97$&$-5.91 $&$1.25$&$-1.40$&$-0.33          $\\
\hline
\multicolumn{13}{c}{${^4}$He}\\
\hline
$0.8    $&$-29.74$&$55.22$&$-192.63$&$15.40$&$82.55$&$-5.23$&$82.06$&$-21.64$&$-37.53$&$2.63$&$ -8.99$&$-1.57          $\\
$1.0    $&$-29.36$&$50.64$&$-171.82$&$ 2.30$&$63.12$&$ 0.10$&$63.30$&$-15.04$&$-18.33$&$9.72$&$-12.11$&$-1.15          $\\
$1.2    $&$-29.19$&$49.15$&$-143.97$&$-2.49$&$46.28$&$ 1.55$&$46.41$&$-13.07$&$-11.33$&$7.94$&$ -8.63$&$-1.04          $\\
\hline
\end{tabular}

\end{table}

For the other three values of the cut-off that we considered, 
the two-body contributions to the ground state energies do indeed decrease
with increasing order of the potential. The ratio between NLO
and LO contributions, as well as that between NNLO and NLO ones, is roughly $1/3$.
This provides a concrete estimate of
our expansion parameter $Q$. Higher-order corrections to our NNLO results
are therefore expected to be of the order of $Q^3\simeq 3\%$. This 
estimate is confirmed by the spread between the values of, for example, 
the binding energy of ${^4}$He for different values of $\sigma$. This
regulator dependence is also at the level of about $3\%$,
cf.~Fig.~\ref{fig:chre4}. The similarity between this dependence and the
expected size of the higher-order contributions suggests that the
higher-order terms are also under control for this range of cut-offs.

The three-nucleon forces, although formally of LO, give rather small
contributions to three- and four-nucleon ground-state energies. This is
consistent with what was observed long ago in phenomenological models 
(see, for example, Ref.~\cite{Wiringa:2002ja}).  In an EFT, however,
the three-body forces serve not only to reproduce the three-body binding
energies, but also to renormalise the latter or, in our approach, to 
cancel the regulator dependence of certain observables. Since the LO and
NLO three-nucleon potentials are fitted to reproduce the ${^3}$H
ground-state energy, one cannot use the latter observable to judge the
regulator dependence. However, one can use other three-body
observables such as the ${^3}$H and ${^3}$He charge radii.
As shown in Fig.~\ref{fig:chr34}, the cut-off dependence of the 
three-body charge radii is numerically smaller than the expected $3\%$
spread, especially in the case of ${^3}$H. This indicates that the wave
functions obtained for the combination of the LO and NLO three-nucleon
forces fixed by fitting the ${^3}$H ground-state energy give consistent
results for other three-body observables.

One can also see a mismatch between the sizes of the LO and NNLO 
three-nucleon potential contributions in Table~\ref{tab:convergence}
(the $B_1$ and $B_2$ terms): the subleading contribution
is sometimes larger than the leading one, especially for the 
shortest-range potential. Nonetheless, the total contribution from these
interactions seems to be well under control.

Finally, the CSB three-nucleon force, Eq.~\eqref{eq:vnnncsb}, which is 
fitted so as to reproduce ${^3}$He binding energy, gives a very small
contribution to the ground-state energies (except in the case of
$\sigma=0.8$~fm). One would also expect this contribution to be larger in
${^4}$He than in ${^3}$He, as the former nucleus is more dense. 
The numbers in Table~\ref{tab:convergence} confirm this expectation.

Our analysis of the contributions at different orders thus suggests a value of
the expansion parameter $Q\simeq 1/3$, which translates to an accuracy
of about $3\%$ for our NNLO calculation. This estimate is
compatible with the apparent mismatch between our result and the 
observed ${^4}$He ground-state energy. It is also consistent with 
the spread between the values of this energy obtained for choices of
the regulator scale $\sigma$, which is also at the level of $3$--$5\%$.
Three-body observables, such as the ${^3}$H and ${^3}$He charge radii, 
show even better agreement with empirical data, which is as expected since the three-nucleon
potential strengths are fitted to the triton ground-state energy. 
CSB interactions, such as the NNLO CSB three-nucleon force, make
rather small contributions, but are important in order to obtain
a satisfactory description of the data.

\end{document}